\documentclass{svmult}

\usepackage{makeidx}         
\usepackage{graphicx}        
\usepackage{multicol}        
\usepackage[bottom]{footmisc}

\makeindex             

\def\be{\begin{equation}}
\def\ee{\end{equation}}
\def\bc{\begin{center}}
\def\ec{\end{center}}
\usepackage[latin1]{inputenc}
\usepackage[english]{babel}

\usepackage{color}
\usepackage{latexsym}
\usepackage{amssymb}
\usepackage{amsfonts}
\usepackage{amsmath}
\usepackage{graphicx}

\begin{document}
\title*{New perspectives in the equilibrium statistical
 mechanics approach to social and economic sciences}
\titlerunning{Statistical
 mechanics approach to social and economic sciences}
\author{Elena Agliari\inst{1} \and Adriano Barra\inst{2} \and  Raffaella Burioni\inst{3} \and Pierluigi Contucci\inst{4} 
}                     
%
%
\institute{Dipartimento di Fisica, Universit\`a di Parma \texttt{elena.agliari@fis.unipr.it}
 \and
Dipartimento di Fisica, Sapienza Universit\`a di Roma  and
Dipartimento di Matematica, Universit\`a di Bologna
\texttt{adriano.barra@roma1.infn.it} \and Dipartimento di Fisica,
Universit\`a di Parma and INFN, Gruppo Collegato di Parma
\texttt{raffaella.burioni@fis.unipr.it} \and Dipartimento di
Matematica, Universit\`a di Bologna \texttt{contucci@dm.unibo.it}
}

\maketitle

\abstract{In this work we review some recent development in the
mathematical modeling of quantitative sociology by means of
statistical mechanics. After a short pedagogical introduction to
static and dynamic properties of many body systems, we develop a
theory for particle (agents) interactions on random graph.

Our approach is based on describing a social network as a graph whose nodes
represent agents and links between two of them stand for a reciprocal
interaction. Each agent has to choose among a dichotomic option
(i.e. agree or disagree) with respect to a given matter and he is
driven by external influences (as media) and peer to peer
interactions. These mimic the
imitative behavior of the collectivity and may possibly be zero if the two nodes
are disconnected.

For this scenario we work out both the dynamics and, given the
validity of the detailed balance, the corresponding equilibria (statics).
Once the two body theory is completely explored, we analyze, on
the same random graph, a {\em diffusive strategic dynamics} with
pairwise interactions, where detailed balance constraint is relaxed.
The dynamic encodes some relevant processes which are expected
to play a crucial role in the approach to equilibrium in social systems,
i.e. diffusion of information and strategic choices. We observe
numerically that such a dynamics reaches a well defined steady
state that fulfills a {\it shift} property: the critical
interaction strength for the canonical phase transition is lower
with respect to the one expected from canonical equilibrium.

Finally, we show how the stationary states of this kind of dynamics
can be described by statistical mechanics equilibria of a diluted
p-spin model, for a suitable non-integer real $p>2$. Several
implications from a sociological perspective are discussed together with
some general outlooks.

} 
\section{Introduction}

Born as a microscopic foundation of thermodynamics, statistical
mechanics provides nowadays a flexible approach to several
scientific problems whose depth and wideness increases
continuously. In the last decades, in fact, statistical mechanics
has invaded fields as diverse as spin glasses \cite{MPV87}, neural
networks \cite{Ami92}, protein folding \cite{Hua07}, immunological
memory \cite{Par90}, and also made some attempt to describe social
networks \cite{CG07}, theoretical economy \cite{Coo05} and urban
planning \cite{CSS00}.

In this paper we study statistical mechanics of imitative
diluted systems, paying particular attention to its
applications in social sciences.

After a review of the statistical mechanics methodology, we introduce the tools, both analytical and numerical, used for the investigation of many-body problems. We apply such a machinery to
study at first the global behavior of a large amount of
dichotomic agents (i.e. able to answer only `yes/no" to a given
question) whose decision making is driven both by a uniform
external influence (as the media) and by pairwise imitative interactions among
agent themselves. In general, the agents making up a community are not all contemporarily in contact with each other, namely, the network representing the social structure is not fully-connected but rather randomly diluted, hence mirroring acquaintances or family relationships.

Even though refined models as small-worlds graphs have recently
been proposed, a standard one for such a network is provided by
the famous Erd\"{o}s-Renyi graph. The model turns out to be non
trivial as, tuning the degree of connectivity and/or the strength
of interaction, a cooperative state among the agents appears.
Moreover, we show the existence of a region of the parameter space
which is more convenient for the global behavior of the society,
i.e. it corresponds to a minimum in the free energy.

As social systems do not need to obey Maxwell-Boltzmann
distribution, detailed balance is not strictly required for their
evolution, so, after having explored the ``canonical" $2$-body
model in full detail, we introduce a more realistic description of
its free temporal evolution by adopting a {\em diffusive strategic
dynamics} \cite{ABCV06a,ABCV06b}. This dynamic takes into account
two crucial aspects, which are expected to be effective in the
temporal evolution of social systems, i.e. diffusion of
information and strategic choices. We implement it on the same
Erd\"{o}s-Renyi graph and study its equilibria. Each agent is
selected through a diffusive rule, and a flip in its dichotomic
status is not weighted ``a la Glauber" \cite{Ami92} but rather
according to a strategic rule which produces the maximum energy
gain. We stress that this operation involves more-than-two-body
effective interactions, as the chosen agent interacts both with
the first selected one as well as with its nearest neighbors, as a
whole. This dynamics is shown to relax to a well defined steady
state, where all the properties of stationarity are recovered
\cite{EM90}, however the  strength of the interactions at the
critical line is lower, of a few percent, than the expected. The
whole scenario suggests a ``latent'' many-body coupling influence,
encoded into the particular rule for selecting the agents. This is
also corroborated by further numerical analysis. As a consequence,
we work out analytically a theory for the randomly diluted
$p$-spin model so to fit an effective $p\in \mathbb{R}$, which
turns out to be $p=2.15$, in order to match the numerical data
available by the dynamics. This result has implication both in
market trends, as well as in quantitative sociology, where the
effective interactions always play an important role in decision
making \cite{Dur99,McF01}.

The paper is organized as follows: In section $2$,
for the sake of completeness, stochastic dynamics for discrete many body problems is outlined, section $3$ deals
with definitions and introduction to their equilibrium via
statistical mechanics. In section $4$, instead, the model we study
is solved in full details with the aim of presenting both a
scenario for these decision makers on random graphs as well as a
general mathematical method which can be extended by the reader to
other models.
\newline
In section \ref{uno} our alternative dynamics is introduced and
shortly discussed; then further numerical investigations toward a
better understanding of a $p>2$ behavior are presented.
\newline
In section \ref{tre} the randomly diluted p-spin model is defined
and exploited in all details, both analytically (within the cavity
field framework) as well as numerically (within a Monte Carlo
approach). Full agreement is found among the two methods. At the end, the last section is left for conclusions: the
effective interaction is found and its implications analyzed.
\newline
Furthermore, even though the paper is written within a theoretical
physics approach, remarks concerning the application to
quantitative sociology are scattered throughout the work.
In particular, in the conclusion,  a toy application of the
outlined theory to trades in markets is shown.

\section{A brief introduction to many-body dynamics}
In this section we introduce the fundamental principles of
stochastic dynamics used to simulate the relaxation to equilibrium
of the systems we are interested in. Even though for discrete
systems two kinds of dynamics are available, parallel and
sequential, we are going to deepen only the latter as is the one
we will implement thought the chapter. Although the topic is well
known (see e.g. \cite{LL80,Lig99}), for the sake of completeness
and in order to offer to the reader a practical approach to these
models, we present the underlying theory.

\subsection{The model}
Let us consider an ensemble of $N$ agents labelled as $i=1,..,N$.
Each agent has two possible choices, with respect to a given
situation, which are encoded into a variable $\sigma_i = \pm 1$,
say $\sigma_i=+1$ is ``agreement" and viceversa for $-1$. Each
agent experiences an external influence (for example by media)
which is taken into account by the one-body coupling
$H_1(\sigma;\theta)$
\begin{equation}\label{eq:H_1}
H_1(\sigma;\theta)= -\sum_{i=1}^N \sigma_i(t) \theta_i(t),
\end{equation}
where $\theta_i(t)$ is the stimulus acting on the $i^{th}$ agent
at a given time $t$ and $\sigma_i(t)$ is the opinion of the
$i^{th}$ agent at the same time.
\newline
The interactions among the other agents are encoded into the
$H_0(\sigma;\textbf{J})$ term as follows
\begin{equation}\label{eq:H_0}
H_0(\sigma;\textbf{J})= -\sum_{i}^{N} \sum_{j>i}^{N}
J_{ij}(\alpha)\sigma_i(t) \sigma_j(t).
\end{equation}
For the moment there is no need to introduce explicitly the
dilution of the underlying random network  as the scheme applies
in full generality and may be a benchmark for future development
by the reader himself. We only stress that $J_{ij}(\alpha)$ is
quenched, i.e. does not evolve with time, and can be thought of as
a symmetric adjacency matrix in such a way that a zero entry
$J_{ij}=0$ means that the agents $i$ and $j$ are not in contact
with each other, viceversa for $J_{ij}=1$ there is a link between
them. The ratio of connections $\sum_{i,j}^N J_{ij}/N^2$ is tuned
by a parameter $\alpha$, such that for $\alpha \to \infty$ the
graph recovers the {\itshape fully connected} one, while for
$\alpha = 0$ the graph is completely disconnected. Overall, the
Hamiltonian defining the model is the sum of the two contributes,
namely (with a little abuse of notation, thinking at $J$ as
$(\textbf{J},\theta)$) $H(\sigma;J)= H_0(\sigma;\textbf{J}) +
H_1(\sigma;\theta)$.
With the signs as they are here, a positive
value of $J_{ij}$ makes the relevant spins want to line up
together, i.e. to share the same opinion, and each spin also wants
to be aligned with the corresponding external field.

The investigation of the properties displayed by systems described
by this kind of Hamiltonian are both analytical and numerical. The
former relies on series expansions, field theoretical methods,
cavity and replica approaches. The latter are mainly based on
Monte Carlo simulations where we directly simulate the temporal
evolution of the system in such a way that an expectation value is
calculated as a time average over the states that the system
passes through. However, it must be underlined that the
Hamiltonian contains no dynamical information, hence we have to
\textit{choose} a dynamic for our simulation, namely a rule for
changing from one state to another during the simulation, which
results in each state appearing with exactly the probability
appropriate to it. Several possibility have been introduced in the
past, ranging from deterministic, e.g. Q2R dynamics, to
stochastic, e.g. Glauber algorithm and Wolff algorithm.

\subsection{Transition rates and Markov process}
In statistical mechanics, Maxwell-Boltzmann statistics (hereafter
simply "Boltzmann statistics") describes the statistical
distribution of material particles over various energy states in
thermal equilibrium, when the temperature is high enough and
density is low enough to render quantum effects negligible. Now,
given the generic configuration $\mathbf{\sigma} = \{ \sigma_i
\}_{i=1,...,N}$ according to Boltzmann statistics,
 the value of a thermodynamic observable
$X(\beta;\mathbf{J})$ is given by
\begin{equation}
\label{eq:canonic_av}
X(\beta;\mathbf{J})=\langle X(\mathbf{\sigma};\mathbf{J}) \rangle_{\beta} =
\frac{\sum_{\{ \sigma_i\}}X(\mathbf{\sigma};\mathbf{J}) e^{-\beta H(\mathbf{\sigma};\mathbf{J})} }{{
\sum_{\{\sigma_i\}}e^{-\beta H(\mathbf{\sigma};\mathbf{J})}}},
\end{equation}
where $\beta$ represents the inverse of the temperature (that sometimes we call ``noise''), i.e.
$\beta \equiv (k_B T)^{-1}$, being $k_B$ the Boltzmann constant,
and the brackets are implicitly defined by the r.h.s. of eq.$(3)$.

Monte Carlo techniques \cite{NB01} work by choosing a subset of
states ${\tilde{S}}$ at random from some probability distribution
$p_{\mathbf{\sigma}}$ which we specify. Our best estimate of the
quantity $X(\mathbf{\sigma};\mathbf{J})$ is then given by the
so-called \textit{estimator} $\langle
X(\mathbf{\sigma};\mathbf{J}) \rangle _{\beta,\tilde{S}}$:

\begin{equation}
\label{eq:canonic_av_red} \langle X(\mathbf{\sigma};\mathbf{J})
\rangle_{\beta,\tilde{S}}= \frac{\sum_{\{ \mathbf{\sigma} \in
\tilde{S}\}} p_{\mathbf{\sigma}}^{-1}
X(\mathbf{\sigma};\mathbf{J}) e^{-\beta
H(\mathbf{\sigma};\mathbf{J})} }{{\sum_{\{ \mathbf{\sigma} \in
\tilde{S}\}} p_{\mathbf{\sigma}}^{-1} e^{-\beta
H(\mathbf{\sigma};\mathbf{J})} }}.
\end{equation}
The estimator has the property that, as the number of sampled
states $|\tilde{S}|$ increases, it becomes a more and more
accurate estimate of $X(\beta;\mathbf{J})$, and, as $|\tilde{S}| \rightarrow
\infty$ we have $\langle X(\mathbf{\sigma};\mathbf{J}) \rangle _{\beta,\tilde{S}} = \langle X(\mathbf{\sigma};\mathbf{J})
\rangle_{\beta}$.

The choice made for $p_{\mathbf{\sigma}}$ is based on the following
argument: when in equilibrium, the system is not sampling all
states in $S$ with equal probability, but according to the
Boltzmann probability distribution. Hence, the strategy is this:
instead of selecting the subspace $\tilde{S}$ in such a way that
every state of the system is as likely to gets chosen as every
other, we select them so that the probability that a particular
state $\textbf{s}$ is chosen is $p_{\mathbf{\sigma}} =
p_{\mathbf{\sigma}}^{(eq)}= Z^{-1} e^{-\beta H(\mathbf{\sigma})}$, $Z$ being a proper normalization factor by now. 

The estimator then simplifies into a simple arithmetic average
\begin{equation}
\label{eq:canonic_av_MC} \langle X(\mathbf{\sigma};\mathbf{J}) \rangle _{\beta,\tilde{S}} =
\frac{1}{|\tilde{S}|} \sum_{\mathbf{\sigma} \in \tilde{S}}
X(\mathbf{\sigma};\mathbf{J}).
\end{equation}
This definition for $\langle X(\mathbf{\sigma};\mathbf{J}) \rangle _{\beta,\tilde{S}}$ works much
better than the one we would obtain from a uniform distribution
for $p_{\mathbf{\sigma}}$, especially when the system is spending the
majority of its time in a small number of states. Indeed, the
latter will be precisely the states sampled most often, and the
relative frequency with which we select them will correspond to
the amount of time the real system would spend in those states.

Therefore, we need to generate an appropriate random set of
states, according to the Boltzmann weight $p_{\mathbf{\sigma}}^{(eq)}$. In
general, Monte Carlo schemes rely on \textit{Markov processes} as
the generating engine for the set of states to be used. Let us
introduce a (normalized) transition probability
$W [\mathbf{\sigma},\mathbf{\sigma}']$ for any pair
$\mathbf{\sigma},\mathbf{\sigma}'$ of configurations in the phase space
$S$. Such a set of transition probabilities, together with the
specification of an initial configuration, allows to construct a
Markov chain of configurations,
$\tilde{S}_{\tau}=(\mathbf{\sigma}_1,\mathbf{\sigma}_2,...,\mathbf{\sigma}_{\tau})$. The
Markov process is chosen in such a way that, when it is run for
long enough, starting from any state of the system, it will
eventually produce a succession of states which appear according
to the canonical distribution. In order to achieve this, two
conditions are sufficient: the condition of \textit{ergodicity}
and of \textit{detailed balance}.

The former is the requirement that it must be possible, for the
Markov process, to reach any state of the system from any other
state, if it is run for long enough. Otherwise stated, $\forall \,
\mathbf{\sigma},\mathbf{\sigma}',\, \exists \, t :
W^t[\mathbf{\sigma},\mathbf{\sigma}']$ is non null, where
$W^t[\mathbf{\sigma},\mathbf{\sigma}']$ just represents the probability of
reaching $\mathbf{\sigma}'$ from $\mathbf{\sigma}$ in $t$ steps. The ergodic condition is also
consistent with the fact that, in the Boltzmann distribution,
every state $\mathbf{\sigma}$ appears with non-zero probability. On the
other hand, notice that this condition does not require that
$\displaystyle W(\mathbf{\sigma},\mathbf{\sigma}')~\neq~0, \,\, \forall \,
\mathbf{\sigma},\mathbf{\sigma}'.$

Conversely, the detailed balance condition ensures that, in the
limit $\tau  \rightarrow \infty$, a given configuration $\mathbf{\sigma'}$
appears in the Markov chain $\tilde{S}_{\tau}$ just according to the
probability distribution $p_{\mathbf{\sigma'}}^{(eq)}$. The detailed balance condition requires that the system is in equilibrium (the rate of transitions into and out of any state must be equal) and that no limit cycles are present. As a result, the detailed balance condition can be stated as
\begin{equation}
\label{eq:detailed_balance}p_{\mathbf{\sigma}}^{(eq)}W[\mathbf{\sigma},\mathbf{\sigma}']
=p_{\mathbf{\sigma}'}^{(eq)} W[\mathbf{\sigma}',\mathbf{\sigma}],
\end{equation}
where the l.h.s. represents the overall rate at which transitions
from $\mathbf{\sigma}$ to $\mathbf{\sigma}'$ occur in the system, while the
r.h.s. is the overall rate for the reverse transition. This
condition makes the system exhibit time-reversal symmetry at each
move and it provides a sufficient (but not necessary) condition
ensuring that the application of these transition probabilities
leads the system to an equilibrium distribution irrespective of
the initial state.
Now, since we wish the equilibrium distribution to be the
Boltzmann one, we choose $p_{\mathbf{\sigma}}^{(eq)}= Z^{-1} e^{-\beta
H(\mathbf{\sigma};\mathbf{J})}$, obtaining
\begin{equation}
\label{eq:detailed_balance2}
W[\mathbf{\sigma},\mathbf{\sigma}']=W[\mathbf{\sigma}',\mathbf{\sigma}]e^{-\beta[H(\mathbf{\sigma}';\mathbf{J})
- H(\mathbf{\sigma};\mathbf{J})]}.
\end{equation}

The constraints introduced so far still leave a good deal of
freedom over the definition of the transition probabilities.
Indeed, the choice of a proper transition probability to apply to
the system under study is crucial. In fact there is no kinetic
information in the Hamiltonian given by eq.~(\ref{eq:H_1}) and by eq.~(\ref{eq:H_0}), as it only contains
information about spin orientation and the spatial distribution of
lattice sites. It is the transition probability which provides a
dynamics, i.e. a rule according to which the system evolves.

In the following, we will be especially interested in the so
called \textit{single-spin-flip} dynamics, which means that the
states involved in the transition only differ for the value of a
single spin variable.
More precisely, in this kind of dynamics, given the configuration $\mathbf{\sigma}$, at each time step a single agent $i \in [1,...,N]$,
is randomly chosen among the $N$ and updated with probability $W[\mathbf{\sigma}, F_i \mathbf{\sigma}]$ to give rise to the the configuration $F_i \mathbf{\sigma}$, where the $N$ spin-flip operators $F_i$
 is defined as
$
F_i \sigma \equiv F_i \{\sigma_1,...,\sigma_i, ..., \sigma_N\} = \{ \sigma_1,...,
-\sigma_i, ..., \sigma_N \}.$

The \textit{Metropolis} and the \textit{Glauber} dynamics
\cite{NB01} are examples of stochastic single-spin-flip dynamics.
In particular, for the latter one has for the transition rates
\begin{equation}
W[\sigma;F_i\sigma] = \Big( 1 + \exp(\beta
\Delta_i H(\mathbf{\sigma}; \mathbf{J})) \Big)^{-1},
\end{equation}
where
$ \Delta_i H(\mathbf{\sigma}; \mathbf{J})
= H(F_i \mathbf{\sigma};  \mathbf{J}) - H( \mathbf{\sigma};  \mathbf{J})$.

\section{Equilibrium behavior}
In synthesis, thermodynamics describes all the macroscopic
features of the system and statistical mechanics allows to obtain
such a macroscopic description starting from its microscopic
foundation, that is, obtaining the global society behavior by
studying  the single agent based dynamics, and then, using
Probability Theory, for averaging over the ensemble with the
weight encoded by the equilibrium distribution
$p_{\mathbf{\sigma}}^{(eq)}$. This scenario is fully derivable
when both the internal energy density of the system
$e(\beta,\alpha)$ and the entropy density $s(\beta,\alpha)$ are
explicitly obtained (we are going to introduce such quantities
hereafter). Then, the two prescription of minimizing the energy
$e(\beta,\alpha)$ (minimum energy principle) and maximizing
entropy $s(\beta,\alpha)$ (second law of thermodynamics) give the
full macroscopic behavior of the system, expressed via suitably
averages of its microscopic element dynamics. To fulfil this task
the free energy $f(\beta,\alpha)= e(\beta,\alpha) -
\beta^{-1}s(\beta,\alpha)$ turns out to be useful because, as it
is straightforward to check, minimizing this quantity corresponds
to both maximizing entropy and minimizing energy (at the given
temperature), furthermore, and this is the key bridge, there is a
deep relation among it and the equilibrium measure
$p_{\mathbf{\sigma}}^{(eq)}$, in fact
\begin{eqnarray}
\nonumber p_{\mathbf{\sigma}}^{(eq)} &\propto& \exp(-\beta H(\sigma;\textbf{J})), \\
f(\beta,\alpha) &=& \lim_{N \to \infty} f_N(\beta,\alpha)=
\lim_{N \to\infty} \frac{-1}{\beta N} \log \sum_{\sigma} \exp(-\beta H_N(\sigma;\textbf{J})), \\
e(\beta,\alpha) &=& \lim_{N \to \infty} e_N(\beta,\alpha) =
\lim_{N \to \infty} -\partial_{\beta} (\beta f_N(\beta,\alpha)), \\
s(\beta,\alpha) &=& \lim_{N \to \infty} s_N(\beta,\alpha) = \lim_{N
\to \infty} \Big( f_N(\beta,\alpha) +
\beta^{-1}\partial_{\beta}(\beta f_N(\beta, \alpha))\Big). 
\end{eqnarray}
So, once explicitly obtained the free energy, equilibrium behavior
is solved and the whole works once the equilibrium probability
distribution is known, provided we use the Hamiltonian $H(\sigma;
J)$.
\newline
Before proceeding in this derivation, we need some preliminary
definitions:
\newline
At first, in the following, it will be convenient to deal with the
pressure $A(\beta,\alpha)$, defined as \be A(\beta,\alpha)=
\lim_{N \to \infty}A_N(\beta,\alpha) = -\beta \lim_{N \to \infty}
f_N(\beta,\alpha); \ee we stress that often we are going to
consider results in the ``thermodynamic limit" $N\to\infty$: such
procedure allows us to use implicitly several theorem of
convergence of random variables from Probability Theory and, when
the number of agents is large enough, the agreement among results
at finite $N$ and results in the thermodynamic limit is excellent,
as usually the two differ by factors $O(N^{-1})$ or at worse
$O(N^{-1/2})$.

Let us now further introduce the partition function defined as
\begin{equation}
Z_N(\beta,\alpha) = \sum_{\sigma_N}e^{-\beta
H_N(\sigma,\textbf{J})} = \sum_{\sigma}p_{\mathbf{\sigma}}^{(eq)}.
\end{equation}
As we do not want a sample-dependent theory, using $\mathbb{E}$
for the average over the quenched variables (i.e. the
connectivity), the quenched pressure can be written as
\begin{equation}\nonumber
A_{N}(\beta,\alpha) = \frac{1}{N}\textbf{E}\ln Z_N(\beta,\alpha),\
\end{equation}
the Boltzmann state is given by
\begin{equation}
\omega(g(\sigma,\textbf{J})) = \frac{1}{Z_{N}(\beta,\alpha)}
\sum_{\sigma_N} g(\sigma;\textbf{J}) e^{-\beta
H_{N}(\sigma;\textbf{J})},
\end{equation}
with its replicated form on $s$ replicas defined as
\begin{equation}
\Omega(g(\sigma;\textbf{J})) = \prod_s \omega^{(s)}(g(\sigma^{(s)};\textbf{J}))
\end{equation}
and the total average $\langle g\rangle$  as
\begin{equation}
\langle g \rangle = \textbf{E}[\Omega(g(\sigma;\textbf{J}))].
\end{equation}

Let us introduce further, as order parameters of the theory, the
multi-overlaps \be
q_{1...n}=\frac1N\sum_{i=1}^{N}\sigma^{(1)}_{i}...\sigma^{(n)}_{i},
\ee with a particular attention to the magnetization $m = q_1
=(1/N)\sum_{i=1}^{N}\sigma_{i}$ and to the two replica overlap
$q_{12} = (1/N)\sum_{i=1}^{N}\sigma_{i}^1\sigma_i^2$.
\newline
It is important to stress that the magnetization, which plays the
role of the principal order parameter (able to recognize the
different macroscopic phases displayed by the system), accounts
for the averaged opinion into the social network, such that if
$\langle m \rangle = 0$ there is no net preference in global
decision, while for $\langle m \rangle \to 1$ there is a sharp
preference toward the ''yes" state and viceversa for $\langle m
\rangle \to -1$. Analogously $ \langle q \rangle$ accounts for
similarity among two different ''replicas" of the system (two
independent realization of the adjacency matrix).
\newline
It is easy to check that when
$\beta \to 0$ (or the interaction strength, that is always
coupled with the noise-), details of the Hamiltonian
are unfelt by the agents, which will be on average one half up
(yes) and one half down (no), giving a null net contribution to
$\langle m \rangle$.
\newline
At the contrary when the system is able to experience the rules
encoded into the Hamiltonian it is easy to see that \be
-\frac{\partial f_N(\beta,a)}{\partial \theta_i} = \langle
\sigma_i \rangle \neq 0, \ \
-\frac{\partial{f_N(\beta,a)}}{\partial J_{ij}} = \langle \sigma_i
\sigma_j \rangle \neq 0. \ee
\newline Averaging over the whole space of choices, we get the
macroscopic response of the system in terms of the magnetization
$\langle m \rangle$.
\newline In the thermodynamic limit, further, self-averaging for
this order parameter is expected to hold, which is expressed via
$$
\lim_{N\to\infty} \langle m_N^2 \rangle = \lim_{N\to \infty}
\langle m_N \rangle^2,
$$
that means that the mean-value of the order parameter is not
affected by the details of the microscopic structure in the $N\to
\infty$ limit (it is an expression of the Central Limit Theorem in
this framework).

From a purely thermodynamical viewpoint the equilibrium behavior
(the phase diagram) is fundamental because it gives both the phase
diagram and the critical scenario, so to say, the regions in the
space of the tunable parameters $\beta, \alpha$ where the model
displays a paramagnetic (independent agent viewpoint) or
ferromagnetic (collective agent viewpoint) behavior, by which
global decision on the whole society can not leave aside.
\newline
To obtain a clear picture of the equilibrium of the social
network, we use standard techniques of statistical mechanics for
positive valued interactions, namely the smooth cavity field
technique \cite{Bar06}.
\newline
For simplicity, as conceptually this does not change the picture,
we deal with the simpler case $\theta_i = \theta \ \forall i \in
[1,N]$.

\section{Equilibrium statistical mechanics of the ``$2$-body" model}

The ``2-body" model has a long history in physics, having
particular importance in interaction theories. In fact, from one
side, (apart historical problems dealing with the deterministic
dynamical evolution of the $3$-body problem), for a long time the
interaction in physics were thought of as particle scattering
processes and in these events the probability of a more-than-$2$
bodies instantaneous interaction were effectively negligible. From
the other side, the structure of the $2$-body energy is quadratic
in its variables and
 this encodes several information: firstly, from a probabilistic
viewpoint, the Maxwell-Boltzmann probability distribution assumes
the form of a Gaussian, then, the forces (the derivatives of the
energy with respect to its variable) are linear, such that
superposition principle and linear response theory do hold and TLC
is respected. However, as we will see later, neither of these
properties of physical systems need to be strictly obeyed in
social theories and the interest to $p>2$ body interactions will
arise.

In this section we consider diluted systems and systematically
develop the interpolating cavity field method \cite{Bar06} and use
it to sketch the derivation of a free energy expansion: the higher
the order of the expansion, the deeper we could go beyond the
ergodic (agent independent) region. Within this framework we
perform a detailed analysis of the scaling of magnetization (and
susceptibility) at the critical line. The critical exponents turn
out to be the same expected for a fully-connected system. Then, we
perform extensive Monte Carlo (MC) simulations for different graph
sizes and bond concentrations and we compare results with theory.
Indeed, also numerically, we provide evidence that the
universality class of this diluted Ising model is independent of
dilution. In fact the critical exponents we measured are
consistent with those pertaining to the Curie-Weiss model, in
agreement with analytical results. The critical line is also well
reproduced.
\newline
The section is organized as follows: after a detailed and technical introduction
of the model, in Section (\ref{cavatappi}) we introduce the cavity
field technique, which constitutes the framework we are going to
use in Section (\ref{laC}) to investigate the free energy of the
system at general values of temperature and dilution. Section
(\ref{critico}) deals with the criticality of the model; there we
find the critical line and the critical behavior of the main order
parameter, i.e. magnetization.  Section (\ref{numerics}) is
devoted to numerical investigations, especially focused on
criticality.

\subsection{Interpolating with the cavity field}\label{cavatappi}

In this section, after a refined introduction of the model, we
introduce further the cavity field technique on the line of
\cite{Bar06}.

Given $N$ points and families $\{i_{\nu},j_{\nu}\}$ of i.i.d
random variables uniformly distributed on these points, the
(random) Hamiltonian of the diluted Curie-Weiss model is defined
on Ising $N$-spin configurations
$\sigma=(\sigma_1,\ldots,\sigma_N)$ through
\begin{equation}\label{ham}
H_{N}(\sigma,\alpha)=-\sum_{\nu=1}^{P_{\alpha N}}
\sigma_{i_{\nu}}\sigma_{j_{\nu}}\ ,
\end{equation}
where $P_{\zeta}$ is a Poisson random variable with mean $\zeta$
 and $\alpha>1/2$ is the connectivity.
\newline
The expectation with respect to all the \emph{quenched} random
variables defined so far will be denoted by $\mathbb{E}$, and is
given by the composition of the Poissonian average with the
uniform one performed over the families $\{i_{\nu}\}$
\begin{equation}
\textbf{E}[\cdot] = E_PE_i[\cdot] = \sum_{k=0}^{\infty}
\frac{e^{-\alpha N}(\alpha
N)^k}{k!N^p}\sum_{i_{\nu}^1....i_{\nu}^p}^{1,N}[\cdot].
\end{equation}

As they will be useful in our derivation, it is worth stressing
the following properties of the Poisson distribution: Let us
consider a function $g:\mathbb{N}\to\mathbb{R}$, and a Poisson
variable $k$ with mean $\alpha N$, whose expectation is denoted by
$\mathbb{E}$.
\newline
It is easy to verify that
\begin{eqnarray}\label{Pp1}
\mathbb{E}[k g(k)] &=& \alpha N \mathbb{E} [g(k-1)] \\ \label{Pp2}
\partial_{\alpha N}\mathbb{E}[g(k)] &=&
\mathbb{E}[g(k+1)-g(k)]\\ \label{Pp3}
\partial^2_{(\alpha N)^2}\mathbb{E}[g(k)] &=&
\mathbb{E}[g(k+2)-2g(k+1)+g(k)].
\end{eqnarray}

Turning to the cavity method, it works by expressing the
Hamiltonian of a system made of $N+1$ spins through the
Hamiltonian of $N$ spins by scaling the connectivity degree $\alpha$
and neglecting vanishing terms in N as follows \be H_{N+1}(\alpha)
= -\sum_{\nu=1}^{P_{\alpha (N+1)}} \sigma_{i_\nu}\sigma_{j_\nu}
\quad \sim \quad -\sum_{\nu=1}^{P_{\tilde{\alpha} N}}
\sigma_{i_\nu}\sigma_{j_\nu} - \sum_{\nu=1}^{P_{2\tilde{\alpha }}}
\sigma_{i_\nu}\sigma_{N+1} \ee such that we can use the more
compact expression
\begin{equation}\label{hsplit}
 H_{N+1}(\alpha) \sim H_{N}(\tilde{\alpha}) +
\hat{H}_{N}(\tilde{\alpha})\sigma_{N+1}
\end{equation}
with \be\label{decompo} \tilde{\alpha} = \frac{N}{N+1}\alpha
\stackrel{N\rightarrow \infty}{\longrightarrow} \alpha, \qquad
\hat{H}_{N}(\tilde{\alpha}) = - \sum_{\nu=1}^{P_{2\tilde{\alpha}}}
\sigma_{i_\nu}. \ee
\medskip
So we see that we can express the Hamiltonian of $N+1$ particles via the one
of $N$ particles, paying two prices: the first is a
rescaling in the connectivity (vanishing in the thermodynamic
limit), the second being an added term, which will be encoded, at
the level of the thermodynamics, by a suitably cavity function as
follows: let us introduce an interpolating parameter $t \in [0,1]$
 and the cavity function $\Psi_N(\tilde{\alpha},t)$ given by
\begin{eqnarray}
\Psi(\tilde{\alpha},\beta;t) &=& \lim_{N\rightarrow \infty
}\Psi_N(\tilde{\alpha}, \beta; t)  \\
\nonumber &=&\lim_{N\rightarrow \infty } \mathbb{E}\Big[\ln
\frac{\sum_{\{\sigma\}} e^{\beta\sum_{\nu=1}^{P_{\tilde{\alpha}
N}} \sigma_{i_\nu}\sigma_{j_\nu} + \beta
\sum_{\nu=1}^{P_{2\tilde{\alpha }t}} \sigma_{i_\nu}}}
{\sum_{\sigma} e^{\beta\sum_{\nu=1}^{P_{\tilde{\alpha} N}}
\sigma_{i_\nu}\sigma_{j_\nu}}}\Big] = \lim_{N\rightarrow \infty }
\mathbb{E}\Big[\ln \frac{Z_{N,t}(\tilde{\alpha},\beta)}
{Z_{N}(\tilde{\alpha},\beta)}\Big].
\end{eqnarray}
The three terms appearing in the decomposition (\ref{hsplit}) give
rise to the structure of the following theorem which we prove by
assuming the existence of the thermodynamic limit. (Actually we
still do not have a rigorous proof of the existence of the
thermodynamic limit but we will provide strong numerical evidences
in Section \ref{numerics}).
\begin{theorem}\label{main}
In the $N\rightarrow \infty$ limit,  the free energy per spin is
allowed to assume the following representation \be A(\alpha,\beta)
= \ln2 - \alpha \frac{\partial A(\alpha,\beta)}{\partial\alpha} +
\Psi(\alpha,\beta;t=1). \ee
\end{theorem}
\medskip
\textbf{Proof}
\newline
Consider the $N+1$ spin partition function $Z_{N+1}(\alpha,\beta)$
and let us split it as suggested by eq. ($\ref{hsplit})$
\begin{eqnarray}\label{zecca}
&Z_{N+1}&(\alpha,\beta) = \sum_{\sigma_{N+1}} e^{-\beta
H_{N+1}(\alpha)} \sim \sum_{\sigma_{N+1}} e^{-\beta
H_{N}(\tilde{\alpha}) -
\beta\hat{H}_{N}(\tilde{\alpha})\sigma_{N+1}}
\\ \nonumber
 &=& \sum_{\sigma_{N+1}}
e^{\beta\sum_{\nu=1}^{P_{\tilde{\alpha} N}}
\sigma_{i_\nu}\sigma_{j_\nu} + \beta
\sum_{\nu=1}^{P_{2\tilde{\alpha }}} \sigma_{i_\nu}\sigma_{N+1}} =
2 \sum_{\sigma_{N}} e^{\beta\sum_{\nu=1}^{P_{\tilde{\alpha}
N}} \sigma_{i_\nu}\sigma_{j_\nu} + \beta
\sum_{\nu=1}^{P_{2\tilde{\alpha }}} \sigma_{i_\nu}}
\end{eqnarray}
\medskip
where the factor two appears because of the sum over the hidden
$\sigma_{N+1}$ variable. Defining a perturbed Boltzmann state
$\tilde{\omega}$ (and  its replica product
$\tilde{\Omega}=\tilde{\omega}\times...\times\tilde{\omega}$) as
$$
\tilde{\omega}(g(\sigma)) =
\frac{\sum_{\{\sigma_{N}\}}g(\sigma)e^{-\beta
H_N(\tilde{\alpha})}} {\sum_{\{\sigma_{N}\}}e^{-\beta
H_N(\tilde{\alpha})}}, \ \ \ \ \ \tilde{\Omega}(g(\sigma)) =
\prod_i\tilde{\omega}^{(i)}(g(\sigma^{(i)}))
$$
where the tilde takes into account the shift in the connectivity
$\alpha \rightarrow \tilde{\alpha}$ and multiplying and dividing
the r.h.s. of eq.(\ref{zecca}) by $Z_N(\tilde{\alpha},\beta)$, we
obtain
\begin{equation}\label{dinox}
Z_{N+1}(\alpha,\beta) = 2 Z_N(\tilde{\alpha},\beta)
\tilde{\omega}(e^{\beta\sum_{\nu=1}^{P_{2\tilde{\alpha }}}}).
\end{equation}
\medskip
Taking now the logarithm of both sides of eq.(\ref{dinox}),
applying the average $\mathbb{E}$ and subtracting the quantity
$\mathbb[\ln Z_{N+1}(\tilde{\alpha},\beta)]$, we get
\begin{equation}\label{latte}
\mathbb{E}[\ln Z_{N+1}(\alpha,\beta)] - \mathbb{E}[\ln
Z_{N+1}(\tilde{\alpha},\beta)] = \ln2 + \mathbb{E}\Big[\ln
\frac{Z_N(\tilde{\alpha},\beta)}{Z_{N+1}(\tilde{\alpha},\beta)}\Big]
+ \Psi_N(\tilde{\alpha},\beta;t=1)
\end{equation}
\medskip
in the large $N$ limit the l.h.s. of eq.(\ref{latte}) becomes
\begin{eqnarray}
\mathbb{E}[\ln Z_{N+1}(\alpha,\beta)] - \mathbb{E}[\ln
Z_{N+1}(\tilde{\alpha},\beta)] &=& \\ \nonumber (\alpha -
\tilde{\alpha}) \frac{\partial}{\partial\alpha}\mathbb{E}[\ln
Z_{N+1}(\alpha,\beta)]
 &=& \alpha \frac{1}{N+1}\frac{\partial}{\partial\alpha}
\mathbb[\ln Z_{N+1}(\alpha,\beta)] = \alpha \frac{\partial
A_{N+1}(\alpha,\beta)} {\partial\alpha}
\end{eqnarray}
\medskip
then by considering the thermodynamic limit the thesis follows.
 $\Box$

\bigskip

Hence, we can express the free energy via an energy-like term and
the cavity function. While it is well known how to deal with the
energy-like \cite{ABC08,Gue95,GT04}, the same can not be stated
for the cavity function, and we want to develop its expansion via
suitably chosen overlap monomials in a spirit close to the
stochastic stability \cite{AC98,CG05,Par00}, such that, at the
end, we will not have the analytical solution for the free energy
in the whole $(\alpha,\beta)$ plane, but we will manage its
expansion above and immediately below the critical line. To see
how the machinery works, let us start by giving some definitions
and proving some simple theorems:
\begin{definition}
We define the t-dependent Boltzmann state $\tilde{\omega}_t$ as
\begin{equation}\label{dante}
\tilde{\omega}_t(g(\sigma)) = \frac{1}{Z_{N,t}(\alpha,\beta)}
\sum_{\{\sigma\}}g(\sigma) e^{\beta\sum_{\nu=1}^{P_{\tilde{\alpha}
N}} \sigma_{i_\nu}\sigma_{j_\nu} + \beta
\sum_{\nu=1}^{P_{2\tilde{\alpha }t}} \sigma_{i_\nu}}
\end{equation}
where $Z_{N,t}(\alpha, \beta)$ extends the classical partition
function in the same spirit of the numerator of eq.(\ref{dante}).
\end{definition}
As we will often deal with several overlap monomials let us divide
them among two big categories:
\begin{definition} We can split the class of monomials of the
order parameters in two families:
\begin{itemize}

\item We define ``filled'' or equivalently ``stochastically
stable'' all the overlap monomials built by an even number of the
same replicas (i.e. $q_{12}^2$,\ $m^2$,\ $q_{12}q_{34}q_{1234}$).

\item We define ``fillable'' or equivalently ``saturable''
all the overlap monomials which are not stochastically stable
(i.e. $q_{12}$,\ $m$,\ $q_{12}q_{34}$)
\end{itemize}
\end{definition}
We are going to show three theorems that will play a guiding role
for our expansion: as this approach has been deeply developed in
similar contexts (as fully connected Ising model \cite{Bar08a} or
fully connected spin glasses \cite{Bar06} or diluted spin glasses
\cite{GT04}, which are the {\em boundary models} of this subject) we will not show all the details of the proofs, but we
sketch them as they are really intuitive. The interested reader
can deepen this point by looking at the original works.
\begin{theorem}\label{ciccia}
For large $N$, setting $t=1$ we have \be
\tilde{\omega}_{N,t}(\sigma_{i_1}\sigma_{i_2}...\sigma_{i_n}) =
\tilde{\omega}_{N+1}(\sigma_{i_1}\sigma_{i_2}...\sigma_{i_n}\sigma_{N+1}^n)
+ O \left( \frac{1}{N} \right )\ee such that in the thermodynamic limit, if
$t=1$, the Boltzmann average of a fillable multi-overlap monomial
turns out to be the Boltzmann average of the corresponding filled
multi-overlap monomial.
\end{theorem}
\begin{theorem}\label{saturabili}
Let $Q_{2n}$ be a fillable monomial of the overlaps (this means
that there exists a multi-overlap $q_{2n}$ such that
$q_{2n}Q_{2n}$ is filled). We have \be
\lim_{N\rightarrow\infty}\lim_{t\rightarrow1} \langle Q_{2n}
\rangle_t = \langle q_{2n}Q_{2n} \rangle \ee  (examples: for $N
\rightarrow \infty$ we get $\langle m_1 \rangle_t \rightarrow
\langle m_1^2 \rangle,\ \langle q_{12} \rangle_t \rightarrow
\langle q_{12}^2 \rangle,\ \langle q_{12}q_{34} \rangle_t
\rightarrow \langle q_{12}q_{34}q_{1234} \rangle$) 
\end{theorem}
\begin{theorem}\label{saturi}
In the $N\rightarrow\infty$ limit the averages
$\langle\cdot\rangle$ of the filled monomials are t-independent in
$\beta$ average.
\end{theorem}
For the proofs of these theorems we refer to \cite{ABC08}.

It is now immediate to understand that the effect of Theorem $(2)$
on a fillable overlap monomial is to multiply it by its missing
part to be filled (Theorem $(3)$), while it has no effect if the
overlap monomial is already filled (Theorem $(4)$) because of the
Ising spins (i.e. $\sigma_{N+1}^{2n} \equiv 1 \ \forall n \in
\mathbb{N}$).

\bigskip

Now the plan is as follows:  We calculate the $t$-streaming of the
$\Psi$ function in order to derive it and then integrate it back
 once we have been able to express it as an expansion in power series of $t$ with
stochastically stable overlaps as coefficients. At the end we free
the perturbed Boltzmann measure by setting $t=1$ and in the
thermodynamic limit we will have the expansion holding with the
correct statistical mechanics weight.
\medskip
\begin{eqnarray}\label{giura} \frac{\partial\Psi(\tilde{\alpha}, \beta, t)}{\partial t}
&=& \frac{\partial} {\partial t}\mathbb{E}[\ln \tilde{\omega}
(e^{\beta \sum_{\nu=1}^{P_{2\tilde{\alpha }t}} \sigma_{i_\nu}})]
\\ \nonumber
= 2\tilde{\alpha}\mathbb{E}[\ln &\tilde{\omega}& (e^{\beta
\sum_{\nu=1}^{P_{2\tilde{\alpha }t}} \sigma_{i_\nu} +
\beta\sigma_{i_0}})] - 2\tilde{\alpha}\mathbb{E}[\ln
\tilde{\omega} (e^{\beta \sum_{\nu=1}^{P_{2\tilde{\alpha }t}}
\sigma_{i_\nu}})] = 2\tilde{\alpha}\mathbb{E}[\ln \tilde{\omega}_t
(e^{\beta\sigma_{i_0}})]
\end{eqnarray}
\medskip
and by the equality $e^{\beta\sigma_{i_0}}=\cosh\beta +
\sigma_{i_0}\sinh\beta$, we can write the r.h.s. of
eq.(\ref{giura}) as
\medskip
$$
\frac{\partial\Psi(\tilde{\alpha}, \beta, t)}{\partial t} =
2\tilde{\alpha}\mathbb{E}[\ln \tilde{\omega}_t (\cosh\beta +
\sigma_{i_0}\sinh\beta )] = 2\tilde{\alpha}\log\cosh\beta -
2\tilde{\alpha}\mathbb{E}[\ln(1 +
\tilde{\omega}_t(\sigma_{i_0})\theta)].
$$
We can expand the function $\log(1 + \tilde{\omega}_t \theta)$ in
powers of $\theta$, obtaining
\begin{equation}\label{sedici}
\frac{\partial\Psi(\tilde{\alpha},t)}{\partial t} =
2\tilde{\alpha}\ln\cosh\beta -
2\tilde{\alpha}\sum_{n=1}^{\infty}\frac{(-1)^n}{n} \theta^n\langle
q_{1,...,n} \rangle_t.
\end{equation}
We learn by looking at eq.(\ref{sedici}) that the derivative of
the cavity function is built by non-stochastically stable overlap
monomials, and their averages do depend on $t$ making their
$t$-integration non trivial (we stress that all the fillable terms
are zero when evaluated at $t=0$ due to the gauge invariance of
the model). We can escape this constraint by iterating them again
and again (and then integrating them back too) because their
derivative, systematically, will develop stochastically stable
terms, which turn out to be independent by the interpolating
parameter and their integration is straightforwardly polynomial.
To this task we introduce the following
\begin{proposition}\label{stream}
Let $F_s$ be a function of s replicas. Then the following
streaming equation holds
\begin{eqnarray}
&& \frac{\partial\langle F_s \rangle_{t,\tilde{\alpha}}}{\partial t}
= 2\tilde{\alpha}\theta [\sum_{a=1}^s\langle F_s \sigma_{i_0}^a
\rangle_{t,\tilde{\alpha}} - s \langle F_s \sigma_{i_0}^{s+1}
\rangle_{t,\tilde{\alpha}}] \quad
\\ \nonumber
&+& \quad 2\tilde{\alpha}\theta^2 [ \sum_{a<b}^{1,s}\langle F_s
\sigma_{i_0}^a\sigma_{i_0}^b \rangle_{t,\tilde{\alpha}} - s
\sum_{a=1}^s\langle F_s \sigma_{i_0}^a\sigma_{i_0}^{s+1}
\rangle_{t,\tilde{\alpha}} + \frac{s(s+1)}{2!}\langle F_s
\sigma_{i_0}^{s+1}\sigma_{i_0}^{s+2} \rangle_{t,\tilde{\alpha}}]
\quad
\\ \nonumber
&+& \quad 2\tilde{\alpha}\theta^3 [\sum_{a<b<c}^{1,s}\langle F_s
\sigma_{i_0}^a\sigma_{i_0}^b\sigma_{i_0}^c
\rangle_{t,\tilde{\alpha}} - s \sum_{a<b}^{1,s}\langle F_s
\sigma_{i_0}^a\sigma_{i_0}^b\sigma_{i_0}^{s+1}
\rangle_{t,\tilde{\alpha}}
\\ \nonumber
&+& \frac{s(s+1)}{2!}\sum_{a=1}^{s}\langle F_s \sigma_{i_0}^a
\sigma_{i_0}^{s+1}\sigma_{i_0}^{s+2}
\rangle_{t,\tilde{\alpha}}\quad + \frac{s(s+1)(s+2)}{3!} \langle
F_s \sigma_{i_0}^{s+1}\sigma_{i_0}^{s+2}\sigma_{i_0}^{s+3}
\rangle_{t,\tilde{\alpha}}] \,
\end{eqnarray}
where we neglected terms $O(\theta^3)$.
\end{proposition}
For a complete proof of the Proposition we refer to \cite{ABC08}.

\subsection{Free energy analysis}\label{laC}

Now that we exploited the machinery we can start applying it to
the free energy. Let us at first work out its streaming with
respect to the plan $(\alpha,\beta)$:

\begin{eqnarray}
&& \frac{\partial A(\alpha,\beta)}{\partial\beta} = -
\frac{\langle H \rangle}{N} = \frac{1}{N}\mathbb{E}
\Big(\frac{1}{Z_N} \sum_{\{\sigma\}} \sum_{\nu=1}^{P_{\alpha N}}
\sigma_{i_\nu}\sigma_{j_\nu} e^{-\beta H_N(\alpha)}\Big)
\\ \nonumber
&&=\sum_{k=1}^{\infty} k \pi(k-1,\alpha N) \frac{\mathbb{E}}{N}
[\omega(\sigma_{i_k}\sigma_{j_k})_k] = \alpha \sum_{k=1}^{\infty}
\pi(k-1,\alpha N) \mathbb{E}
\Big[\frac{\omega(\sigma_{i_k}\sigma_{j_k}e^{\beta\sigma_{i_k}\sigma_{j_k}})_{k-1}}
{\omega(e^{\beta\sigma_{i_k}\sigma_{j_k}})_{k-1}}\Big]
\\ \nonumber
&&= \alpha \mathbb{E}
\Big[\frac{\omega(\sigma_{i_k}\sigma_{j_k}(\cosh\beta +
\sigma_{i_k}\sigma_{j_k}\sinh\beta))} {\omega(\cosh\beta +
\sigma_{i_k}\sigma_{j_k}\sinh\beta)}\Big] = \alpha
\mathbb{E}\Big[\frac{\omega(\sigma_{i_k}\sigma_{j_k}) + \theta} {1
+ \omega(\sigma_{i_k}\sigma_{j_k})\theta}\Big]
\end{eqnarray}
\medskip
by which we get (and with similar calculations for
$\partial_{\alpha}A(\alpha,\beta)$ that we omit for the sake of
simplicity):
\begin{eqnarray}
\frac{\partial A(\alpha,\beta)}{\partial\beta} &=& \alpha\theta -
\alpha \sum_{n=1}^{\infty} (-1)^n (1 - \theta^2) \theta^{n-1}
\langle q_{1,..,n}^2 \rangle
\\ \label{dAda} \frac{\partial A(\alpha,\beta)}{\partial\alpha} &=& \ln
\cosh\beta - \sum_{n=1}^{\infty} \frac{(-1)^n}{n} \theta^{n}
\langle q_{1,..,n}^2 \rangle
\end{eqnarray}
\bigskip
Now remembering Theorem (\ref{main}) and assuming critical
behavior (that we will verify a fortiori in sec. (\ref{critico}))
we move for a different formulation of the free energy by
considering the cavity function as the integral of its derivative.
In a nutshell the idea is as follows: Due to the second order
nature of the phase transition for this model (i.e. criticality
that so far is assumed) we can expand the free energy in terms of
the whole series of order parameters. Of course it is impossible
to manage all these infinite overlap correlation functions to get
a full solution of the model in the whole ($\alpha,\beta$) plane
but it is possible to show by rigorous bounds that close to the
critical line (that we are going to find soon) higher order
overlaps scale with higher order critical exponents so we are
allowed to neglect them close to this line and we can investigate
deeply criticality, which is the topic of the section.
\newline
To this task let us expand the cavity functions as
\medskip
\begin{eqnarray}
\Psi(\tilde{\alpha},\beta,t) &=& \int_0^t \frac{\partial
\Psi}{\partial t'}dt'
\\ \nonumber
&=& 2\tilde{\alpha}t \log\cosh\beta + \tilde{\beta}\int_0^t\langle
m \rangle_{t',\tilde{\alpha}}dt' -
\frac{1}{2}\tilde{\beta}\theta\int_0^t\langle q_{12}
\rangle_{t',\tilde{\alpha}}dt' + O(\theta^3)
\end{eqnarray}
where $\tilde{\beta} = 2\tilde{\alpha}\theta \rightarrow \beta' =
2\alpha\theta$ for $N\rightarrow \infty$. Now using the streaming
equation as dictated by Proposition $(1)$ we can write the
overlaps appearing in the expression of $\Psi$ as polynomials of
higher order filled overlaps so to obtain a straightforward
polynomial back-integration for the $\Psi$ as they no longer will
depend on the interpolating parameter $t$ thanks to Theorem $(1)$.
\newline
For the sake of simplicity the $\tilde{\alpha}$-dependence of the
overlaps will be omitted keeping in mind that our results are all
taken in the thermodynamic limit and so we can quietly exchange
$\tilde{\alpha}$ with $\alpha$ in these passages.
\medskip
The first equation we deal with is:
\begin{equation}
\frac{d\langle m \rangle_t}{dt} = \tilde{\beta}[\langle m^2
\rangle - \langle m_1m_2 \rangle_t]
\end{equation}
\medskip
where $\langle m_1m_2 \rangle $ is not filled and so we have to go
further in the procedure and derive it in order to obtain filled
monomials:
\begin{eqnarray}\label{m1m2}
&& \frac{d\langle m_1m_2 \rangle_t}{dt} = \\
\nonumber
& = & 2\tilde{\beta}[\langle
m_1^2m_2 \rangle_t - \langle m_1m_2m_3 \rangle_t] +
\tilde{\beta}\theta [\langle m_1m_2q_{12} \rangle - 4\langle
m_1m_2q_{13} \rangle_t + 3\langle m_1m_2q_{34} \rangle_t].
\end{eqnarray}
In this expression we stress the presence of the filled overlap
$\langle m_1m_2q_{12} \rangle$ and of $\langle m_1^2m_2 \rangle_t$
which can be saturated in just one derivation. Wishing to have an
expansion for $\langle m \rangle_t$  up to the third order in
$\theta$, it is easy to check that the saturation of the other
overlaps in the last derivative would carry terms of higher order
and so we can stop the procedure at the next step
\begin{equation}
\frac{d\langle m_1^2m_2 \rangle_t}{dt} = \tilde{\beta}[\langle
m_1^2m_2^2 \rangle] + \tilde{\beta}[\mbox{unfilled terms}] +
O(\theta^2)
\end{equation}
from which integrating back in $t$
\begin{equation}
\langle m_1^2m_2 \rangle_t = \tilde{\beta}[\langle m_1^2m_2^2
\rangle]t
\end{equation}
inserting now this result in the expression (\ref{m1m2}) and
integrating again in $t$ we find
\begin{equation}
\langle m_1m_2 \rangle_t = \tilde{\beta}\theta\langle m_1m_2q_{12}
\rangle t + \tilde{\beta}^2\langle m_1^2m_2^2 \rangle t^2
\end{equation}
and coming back to $\langle m \rangle_t$ we get
\medskip
\begin{equation}\label{magneto}
\quad \langle m \rangle_t = \tilde{\beta}\langle m^2 \rangle t -
\frac{\tilde{\beta}^2\theta}{2}\langle m_1m_2q_{12} \rangle t^2 -
\frac{\tilde{\beta}^3}{3}\langle m_1^2m_2^2 \rangle t^3
\end{equation}
which is the attempted result. Let us move our attention to
$\langle q_{12}\rangle_t$, analogously we can write
\begin{equation}
\frac{d\langle q_{12} \rangle_t}{dt} = 2\tilde{\beta}[\langle
m_1q_{12} \rangle_t - \langle m_3q_{12} \rangle_t] +
\tilde{\beta}\theta [\langle q_{12}^2 \rangle - 4\langle
q_{12}q_{13} \rangle_t + 3\langle q_{12}q_{34} \rangle_t]
\end{equation}
and consequently obtain
\begin{equation}\label{overlap}
\quad \langle q_{12} \rangle_t = \tilde{\beta}\theta \langle
q_{12}^2 \rangle t + \tilde{\beta}^2\langle m_1m_2q_{12} \rangle
t^2 + O(\theta^4).
\end{equation}
With the two expansion above, in the $N\rightarrow\infty$ limit,
putting $t=1$ and neglecting terms of order higher than
$\theta^4$, we have
\medskip
\begin{equation}
\Psi(\alpha,\beta,t=1) = 2\alpha \ln\cosh\beta +
\frac{\beta'}{2}\langle m^2 \rangle - \frac{\beta'^4}{12}\langle
m_1^2m_2^2 \rangle - \frac{\beta'^2\theta^2}{4}\langle q_{12}^2
\rangle - \frac{\beta'^3\theta}{3}\langle m_1m_2q_{12} \rangle
\end{equation}
\medskip
At this point we have all the ingredients to write down the
polynomial expansion for the free energy function as stated in the
next:
\medskip
\begin{proposition}
A general expansion via stochastically stable terms for the free
energy of the random two body interacting imitative agent model
can be written as
\begin{eqnarray}\label{A}
A(\alpha,\beta) &=& \ln2 + \alpha \ln\cosh\beta  +
\frac{\beta'}{2}\left(\beta' - 1\right)\langle m_1^2 \rangle  +
\\ \nonumber
&-&  \frac{\beta'^4}{12}\langle m_1^2m_2^2 \rangle  -
\frac{\beta'^2}{8\alpha}\left(\frac{\beta'^2}{2\alpha} -
1\right)\langle q_{12}^2 \rangle  -
\frac{\beta'^4}{6\alpha}\langle m_1m_2q_{12} \rangle  +
O(\theta^6).
\end{eqnarray}
\end{proposition}
It is immediate to check that the above expression, in the ergodic
region where the averages of all the order parameters vanish,
reduces to the well known high-temperature (or high connectivity)
solution \cite{ABC08} (i.e. $A(\alpha,\beta)=\ln 2 + \alpha
\log\cosh \beta$).
\newline
Of course we are neglecting $\theta^6$ higher order terms because
we are interested in an expansion holding close to the critical
line, but we are not allowed to truncate the series for a general
point in the phase space far beyond the ergodic region.

\subsection{Critical behavior and phase transition}\label{critico}

Now we want to analyze the critical behavior of the model: we find
the critical line where the ergodicity breaks, we obtain the
critical exponent of the magnetization and the susceptibility
$\chi$, which is defined as $\langle \chi \rangle = \beta N
[\langle m^2 \rangle - \langle m \rangle^2]$.

Let us firstly define the rescaled magnetization $\xi_N$ as $\xi_N
= \sqrt{N}m_N$. By applying the gauge transformation $\sigma_i
\rightarrow \sigma_i\sigma_{N+1}$ in the expression for the
quenched average of the magnetization (eq. (\ref{magneto})) and
multiplying it times $N$ so to switch to $\xi^2_N$, setting $t=1$
and sending $N \rightarrow \infty$ we obtain
\medskip
\begin{equation}\label{m2}
\langle \xi_1^2 \rangle = \frac{\beta'^3}{3(\beta'-1)}\langle
\xi_1\xi_2 m_1 m_2 \rangle +
\frac{\beta'^2\theta}{2(\beta'-1)}\langle \xi_1\xi_2 q_{12}
\rangle + O(\frac{\theta^5}{\beta'-1})
\end{equation}
\medskip
by which we see (again remembering criticality and so forgetting
higher order terms) that the only possible divergence of the
(centered and rescaled) fluctuations of the magnetization happens
at the value $\beta' = 1$ which gives $2\alpha\theta=1$ as the
critical line, in perfect agreement with the expected Landau-like
behavior.
\medskip
The same critical line can be found more easily by simply looking
at the expression (\ref{A}) as follows: remembering  that in the
ergodic phase the minimum of the free energy corresponds to a zero
order parameter (i.e.$\sqrt{\langle m^2 \rangle}=0$), this implies
the coefficient of the second order $a(\beta') =
\frac{\beta'}{2}(\beta' - 1)$ to be positive. Anyway immediately
below the critical line values of the magnetization different from
zero must be allowed (by definition otherwise we were not crossing
a critical line) and this can be possible if and only if
$a(\beta') \leq 0$. Consequently (and using once more the second
order nature of the transition) on the critical line we must have
$a(\beta') = 0$ and this gives again $2\alpha\theta = 1$.

\begin{figure}[tb]\bc
\includegraphics[angle=0,width=58mm,height=55mm]{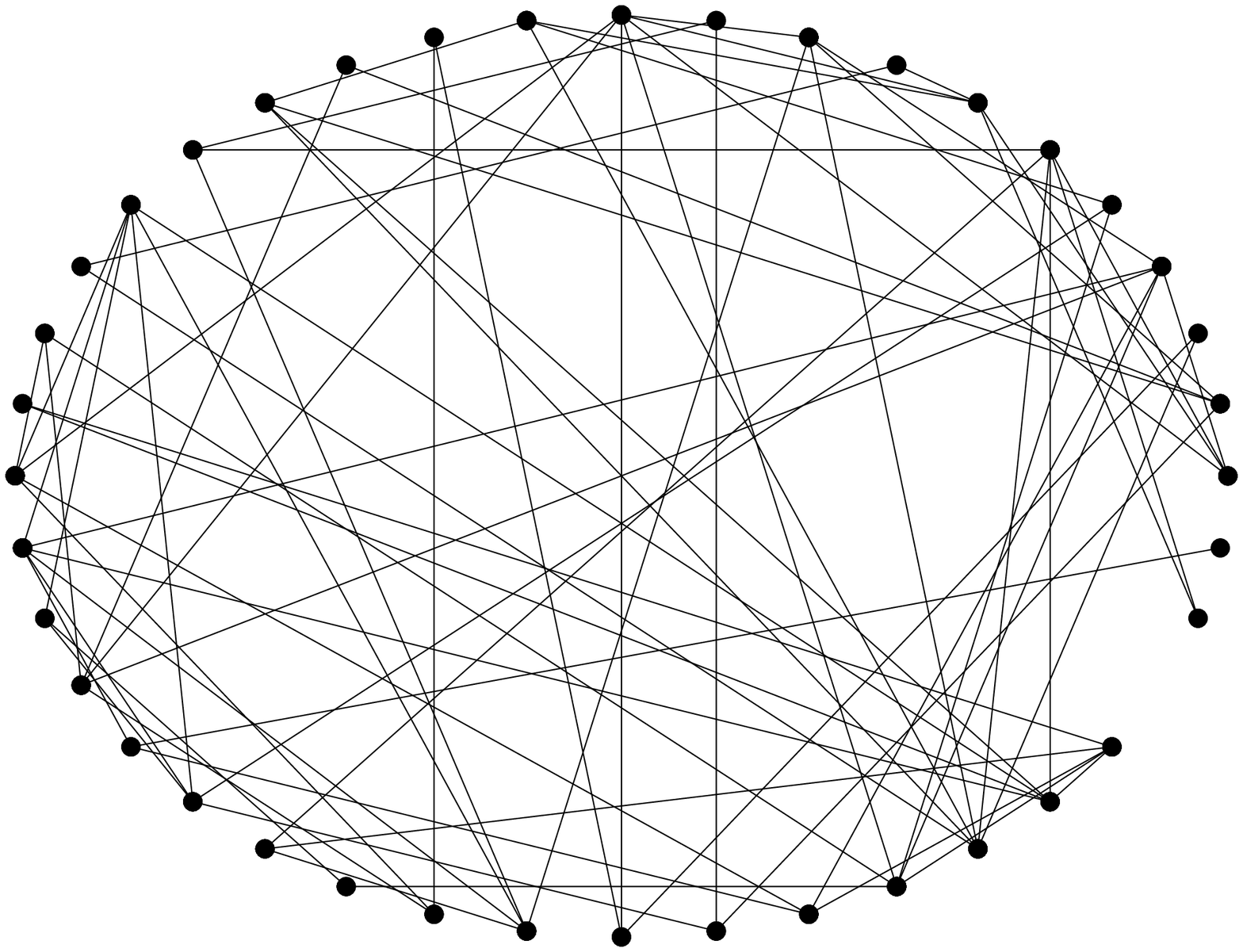}
\includegraphics[angle=0,width=58mm,height=55mm]
{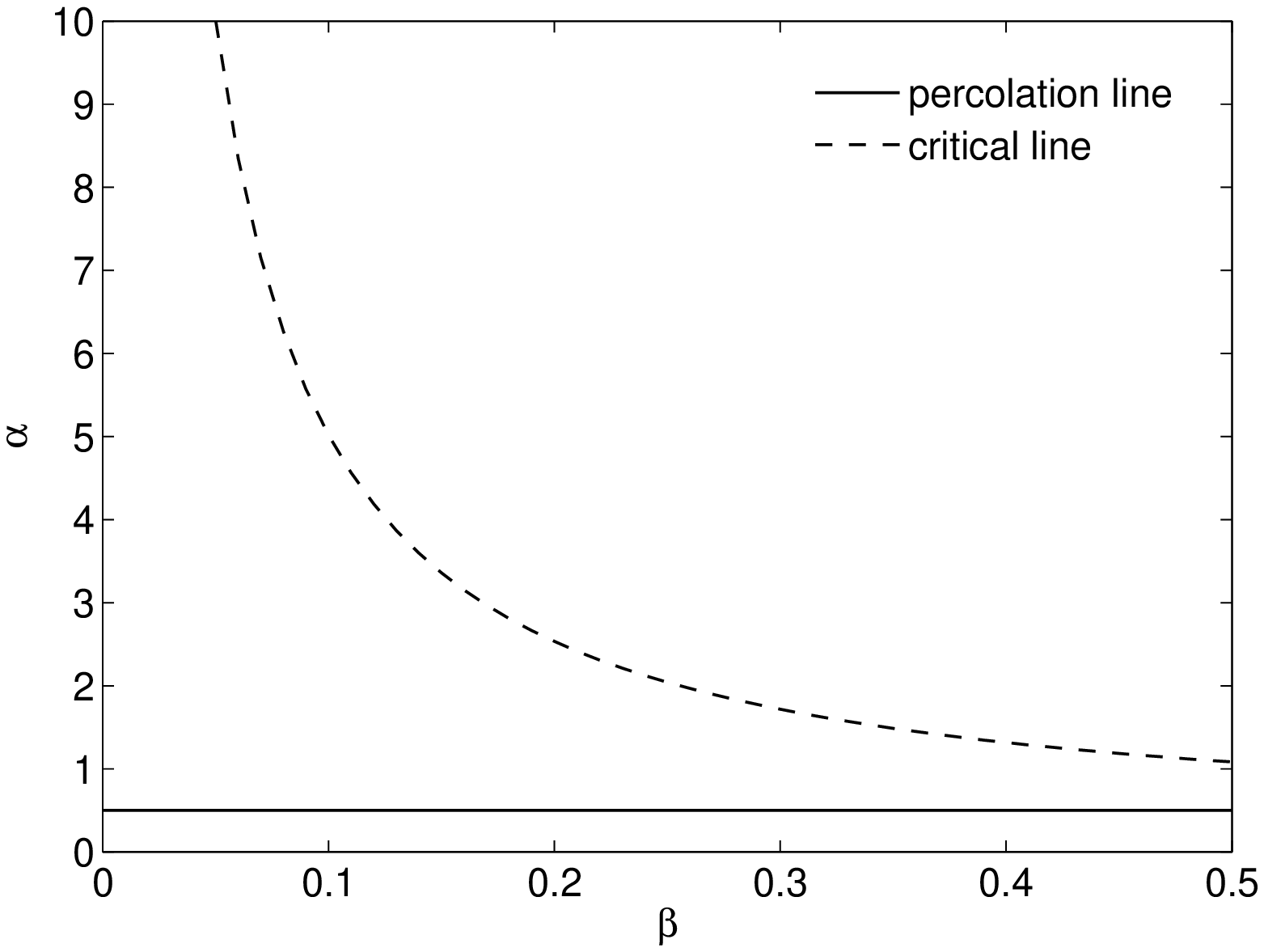} \caption{\label{phase} Left panel: Example of
Erd\"{o}s-Renyi random graph made up of $40$ nodes and with
average degree equal to $4$. Right panel: Phase diagram: below
$\alpha_c=0.5$ there is no giant component in the graph,
$\alpha_c$ defines the percolation threshold. Above $\alpha_c$ at
left of the critical line the system behaves ergodically,
conversely on the right ergodicity is broken and the system
displays magnetization.} \ec
\end{figure}

Now let us move to the critical exponents. Critical exponents are
needed to characterize singularities of the observables at the
critical line and, for us, these indexes are the ones related to
the magnetization $\langle m \rangle$ and to the susceptibility
$\langle \chi \rangle$.
\newline
We define $\tau=(2\alpha \tanh \beta-1)$ and we write $\langle
m(\tau) \rangle \sim G_0 \cdot \tau^{\delta}$ and $\langle
\chi(\tau) \rangle \sim G_0 \cdot \tau^{\gamma}$, where the symbol
$\sim$ has the meaning that the term at the second member is the
dominant but there are corrections of higher order.
\newline
Remembering the expansion of the squared magnetization that we
rewrite for completeness
\begin{equation}\label{m2}
\langle m^2 \rangle = \frac{\beta'^3}{3(\beta'-1)}\langle
m_1^2m_2^2 \rangle + \frac{\beta'^2\theta}{2(\beta'-1)}\langle
m_1m_2q_{12} \rangle + O(\frac{\theta^5}{\beta'-1})
\end{equation}
and considering that using the same gauge transformation $\sigma_i
\rightarrow \sigma_i\sigma_{N+1}$ on (eq.(\ref{overlap})) we have
for the two replica overlap the following representation
\begin{equation}
\langle q_{12}^2 \rangle = - \frac{\beta'^2}{(\beta'\theta -
1)}\langle m_1m_2q_{12} \rangle + O(\theta^6)
\end{equation}
we can arrive by simple algebraic calculations to write down the
free energy, of course close to the critical line, depending only
by the two parameters $\langle m^2 \rangle$ and $\langle q_{12}^2
\rangle$
\medskip
\begin{equation}\label{38}
A(\alpha,\beta) = \ln2 + \alpha \ln \cosh\beta +
\frac{\beta'}{4}\left(\beta' - 1\right)\langle m_1^2 \rangle -
\frac{\beta'^2}{48\alpha}\left(\frac{\beta'^2}{2\alpha} -
1\right)\langle q_{12}^2 \rangle + O(\theta^6).
\end{equation}
\medskip
\newline
By a comparison of the formula obtained by deriving
$A(\alpha,\beta)$ as expressed by eq.(\ref{38}) and the expression
we have previously found (eq. $41$), it is immediate to see that
we have
\medskip
\begin{equation}
\frac{\partial}{\partial\alpha}\Big[\frac{\beta'}{4}(\beta' -
1)\langle m_1^2 \rangle\Big]= \theta \langle m_1^2 \rangle.
\end{equation}
\medskip
Close to the value $\beta'=1$,  making a change of variable $\tau
= \beta' - 1$ with $\partial_{\alpha} = 2\theta
\partial_{\tau}$, we get
\begin{equation}
\frac{\partial}{\partial\alpha}\Big[\frac{\beta'}{4}(\beta' -
1)\langle m_1^2 \rangle\Big] \sim
\frac{\theta}{2}\frac{\partial}{\partial\tau}[\tau\langle m_1^2
\rangle] = \frac{\theta}{2}\langle m_1^2 \rangle +
\frac{\theta\tau}{2}\frac{\partial\langle m_1^2 \rangle}{\partial
\tau} = \theta \langle m_1^2 \rangle,
\end{equation}
by which we easily obtain
\begin{equation}
\frac{\partial\langle m_1^2 \rangle}{\langle m_1^2 \rangle} =
\frac{\partial \tau}{\tau} \qquad \Rightarrow \langle m_1^2
\rangle \sim \tau \qquad \Rightarrow \sqrt{\langle m_1^2 \rangle}
\sim \tau^{\frac{1}{2}} = \tau^{\delta}
\end{equation}
Therefore we get  the critical exponent for the magnetization,
$\delta = 1/2$, which turns out to be the same as in the fully
connected counterpart \cite{Bar08a}, in agreement with the
disordered extension of this model \cite{Bar06}.
\newline
Again, by simple direct calculations, once we get the critical
exponent for the magnetization it is straightforward to show that
the susceptibility $\langle \chi \rangle$ obeys \be \langle \chi
\rangle \sim |\tau|^{-1} = \tau^{\gamma}\ee close to the critical line, by which
we find its critical exponent to be once again in agreement with
the classical fully connected counterpart \cite{Ami92}, i.e. $\gamma=-1$.

\subsection{Numerics}\label{numerics}

\begin{figure}[tb]\bc
\includegraphics[height=60mm,width=78mm]{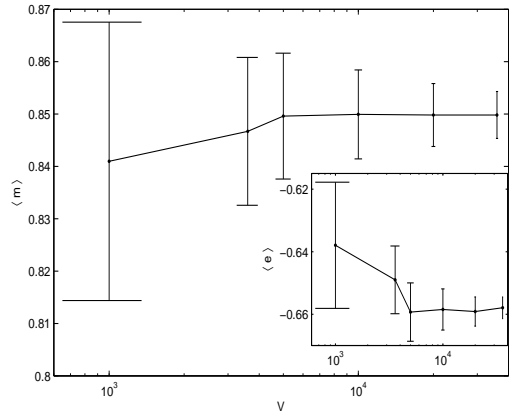}
\caption{\label{fig:fss}Finite size scaling for the  magnetization
 and the internal energy (inset) for $\bar{\alpha}=10$ and
$\frac{\beta}{\bar{\alpha}}=1.67$. All the measurements were
carried out in the stationary regime and the error bars represent
the fluctuations about the average values. We find good indication
of the convergence of the quantities on the size of the system and
thus of the existence of the thermodynamic limit.} \ec
\end{figure}

In this section,  by performing extensive Monte Carlo simulations
with the Glauber algorithm \cite{NB01}, we analyze, from the
numerical point of view, the $2$-body diluted imitative system
previously introduced.
\newline
The Erd\"{o}s-Renyi random graph is constructed by taking $N$
sites and introducing a bond between each pair of sites with
probability $p=\bar{\alpha}/(N-1)$, in such a way that the average
coordination number per node is just $\bar{\alpha}$. Clearly, when
$p=1$ the complete graph is recovered.
\newline
The simplest version of the diluted Curie-Weiss Hamiltonian has a
Poisson variable per bond as
$H_N=-\sum_{ij}\sum_{\nu=0}^{P_{\bar{\alpha}/
N}}\sigma_{i_{\nu}}\sigma_{j_{\nu}}$, and it is the easiest
approach when dealing with numerics.
\newline
For the analytical investigation we choose a slightly changed
version (see eq.(\ref{ham})): each link gets a bond with probability
close to $\alpha/N$ for large $N$; the probabilities of getting
two, three bonds scale as $1/N^2,1/N^3$ therefore negligible  in
the thermodynamic limit.
\newline
Working with directed links (as we do in the analytical framework)
the probability of having a bond on any undirected link is twice
the probability for directed link (i.e. $2\alpha/N$). Hence, for
large $N$, each site has average connectivity $2\alpha$. Finally
in this way we allow self-loop but they add just
$\sigma$-independent constant to $H_N$ and are irrelevant, but we
take the advantage of dealing with an $H_N$ which is the sum of
independent identically distributed random variables, that is
useful for analytical investigation.
\newline
When comparing with numerics consequently we must keep in mind
that $\bar{\alpha} = 2 \alpha$.
\newline
In the simulation, once the network has been diluted, we place a
spin $\sigma_i$ on each node $i$ and allow it to interact with its
nearest-neighbors. Once the external parameter $\beta$ is fixed,
the system is driven by the single-spin dynamics and it eventually
relaxes to a stationary state characterized by well-defined
properties. More precisely, after a suitable time lapse $t_0$  and
for sufficiently large systems, measurements of a (specific)
physical observable $x(\sigma,\bar{\alpha},\beta)$ fluctuate
around an average value only depending on the external parameters
$\beta^{-1}$ and $\bar{\alpha}$.
\newline
Moreover, for a system $(\bar{\alpha},\beta)$ of a given finite
size $N$ the extent of such fluctuations scales as
$N^{-\frac{1}{2}}$ with the size of the system. The estimate of
the thermodynamic observables $\langle x \rangle$ is therefore
obtained as an average over a suitable number of (uncorrelated)
measurements performed when the system is reasonably close to the
equilibrium regime.
\newline
The estimate is further improved by averaging over different
realizations of the same system
$(\bar{\alpha},\beta)$. 
In summary,
$$
\langle x(\sigma,\bar{\alpha},\beta) \rangle = \mathbb{E} \left[
\frac{1}{M} \sum_{n=1}^{M} x(\sigma(t_n))\right] , \; \;
t_n=t_0+n\mathcal{T}
$$
where $\sigma(t)$ denotes the configuration of the magnetic system
at time step $t$ and $\mathcal{T}$ is the decorrelation parameter
(i.e. the time, in units of spin flips, needed to decorrelate a
given magnetic arrangement). In general, statistical errors during a MC run in a given sample
result to be significantly smaller than those arising from the
ensemble averaging.
Figure (\ref{fig:fss}) shows the dependence of the macroscopic
observables $\langle m \rangle$ and $\langle e \rangle$ from the
size of the system; values are obtained starting from a
ferromagnetic arrangement, at the normalized inverse temperature
$\beta/\bar{\alpha}=1.67$. Notice that at this temperature the
system composed of $N=10^4$ is already very close to the
asymptotic regime. Analogous results are found for different
systems $(\bar{\alpha},\beta)$, with $\beta$ far enough from
 $\beta_c$.

\begin{figure}[tb]
\bc
\includegraphics[height=60mm]{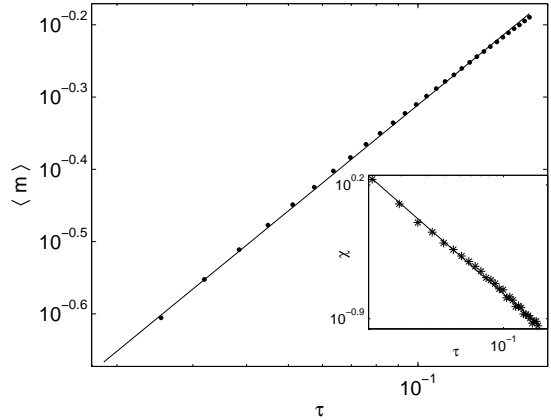}
\caption{\label{fig:crit} Log-log scale plot of magnetization
(main figure) and susceptibility (inset) versus the reduced
temperature $\tau=(|\beta-\beta_c|/\beta_c)^{-1}$ for
$\bar{\alpha} = 10$. Symbols represents data from numerical
simulations performed on systems of size $N=36000$, while lines
represent the best fit.
} \ec
\end{figure}

In the following we focus on systems of sufficiently large size so
to discard finite size effects. For a wide range of temperatures
and dilutions, we measure the average magnetization $\langle m
\rangle$ and energy $\langle e \rangle$, as well as the magnetic
susceptibility $\langle \chi \rangle$

Their profiles display the typical behavior expected for a
ferromagnet (i.e. imitative coupling) and, consistently with the
theory, highlight a phase transition at well defined temperatures
$\beta_c(\alpha)$.
\newline
Now, we investigate in more detail the critical behavior of the
system. We collect accurate data of magnetization and
susceptibility, for different values of $\bar{\alpha}$ and for
temperatures approaching the critical one.
These data are used to estimate both the critical temperature and
the critical exponents for the magnetization and susceptibility.
In Fig.~(\ref{fig:crit}) we show data as a function of the reduced
temperature $\tau = (|\beta-\beta_c|/\beta_c)^{-1}$ for
$\bar{\alpha}=10$ and $\bar{\alpha}=20$. The best fit for
observables is the power law
\begin{eqnarray}
\label{eq:powerlaw}
\langle m \rangle & \sim & \tau^{\delta},\ \beta > \beta_c \\
\langle \chi \rangle  & \sim & \tau^{\gamma}.
\end{eqnarray}
We obtain estimates for $\beta_c(\bar{\alpha})$,
$\delta(\bar{\alpha})$ and $\gamma(\bar{\alpha})$ by means of a
fitting procedure. Results are gathered in Tab. $1$.
\begin{table}[htbp]\label{tabexp}
\begin{center}
\begin{tabular}{p{0.5cm}p{1cm}|p{0.5cm}p{1.2cm}p{1.2cm}p{1.2cm}}
\hline
& $\bar{\alpha}$ & & $\;\beta^{-1}_c$ &  $\;\delta$ & $\;\gamma$ \\
\hline
& 10 && $9.93$    &  $0.48$ & $-0.97$ \\
& 20 && $19.92$   &  $0.49$ & $-1.04$ \\
& 30 && $29.98$   &  $0.48$ & $-1.04$ \\
& 40 && $39.59$   &  $0.50$ & $-1.02$ \\
\hline
\end{tabular}
\end{center}
\caption{Estimates for the critical temperature and the critical
exponents $\delta$ and $\gamma$ obtained by a fitting procedure on
data from numerical simulations concerning Ising systems of size
$N=36000$ and different dilutions (we stress that analytically we
 get $\delta = 0.5$ and $\gamma=-1$). Errors on temperatures are $<2\%$, while for exponents are
within $5 \%$.}
\end{table}
Within the errors ($\leq 2\%$ for $\beta_c$ and $ \leq 5 \%$ for
the exponents), estimates for different values of $\bar{\alpha}$
agree and they are also consistent with the analytical results
exposed in Sec. (\ref{critico})
\newline
We also checked the critical line for the ergodicity breaking,
again finding optimal agreement with the criticality investigated
by means of analytical tools.

\section{Beyond detailed balance: Diffusive strategic dynamics}\label{uno}
As we saw in Section $2$, a dynamics which obeys detailed balance
is constrained to reach the Maxwell-Boltzmann equilibrium. In social context however,
detailed balance actually lacks a clear interpretation, or a data comparison.
In this section we figure out a possible and more realistic dynamics
which aims to mimic opinion spreading and strategic choices on random networks,
without any a priori constraint based on detailed balance.

\subsection{The algorithm}

In order to simulate the dynamical evolution of Ising-like system
as our one several different algorithms have been introduced. In
particular, a well established one is the so-called single-flip
algorithm, which makes the system evolve by means of successive
spin-flips, where we call ``flip'' on the node $j$ the
transformation ${\sigma}_j \rightarrow - {\sigma}_j$ \cite{Lig99}.

More precisely, the generic single-flip algorithm is made up of
two parts: first we need a rule according to which we select a
spin to be updated, then we need a probability distribution which
states how likely the spin-flip is. As for the latter, following
the Glauber rule, given a configuration $\mathbf{\sigma}$, the
probability for the spin-flip on the $j$-th node reads off as
\begin{equation} \label{eq:Glauber}
p(\mathbf{\sigma},j,\mathbf{J}) = \frac{1}{1 + e^{\displaystyle
\beta\Delta H(\mathbf{\sigma},j,\mathbf{J})  }},
\end{equation}
where $\Delta H(\mathbf{\sigma},j,\mathbf{J})= 2\sigma_i\sum_j
J_{ij}\sigma_j$ is the variation in the cost function due to the
flip $\sigma_j \rightarrow - \sigma_j.$ Hence, for single-flip dynamics the
cost variation $\Delta H$, consequent to a flip, only depends on
the spin of a few sites, viz. the $j$-th one undergoing the
flipping process and its $\alpha_j$ nearest-neighbors.
\newline
As for the selection rule according to which sites are extracted,
there exist several different choices, ranging from purely random
to deterministic. In several contexts (condensed-matter physics
\cite{BBCV02}, sociology \cite{ABC09}, etc.) unless no peculiar
mechanisms or strategies are at work, the random updating  seems
to be the most plausible.

In a social context a spin-flip can occur as a result of a direct
interaction (phone call, mail exchange, etc.) between two
neighbors and if agent $i$ has just undergone an opinion-flip he
will, in turn, prompt one out of his $\alpha_i$ neighbors to
change opinion where, we recall, in social context opinion plays the role of the
spin orientation in material systems.

These aspects are neglected by traditional dynamics and can not be
described by a random updating rule.
A different relaxation dynamics, introduced and developed in \cite{ABCV05,ABCV06a,ABCV06b}, is able to take into account these aspects, namely:\\
\textit{i}. the selection rule exhibits a \textit{diffusive
character}: The sequence of sites selected for the updating can be
thought of as the  path of a random walk moving on the
underlying structure.\\ \textit{ii}. the diffusion is
\textit{biased}: The $\alpha_i$ neighbors are not equally likely
to be chosen but, amongst the $\alpha_i$ neighbors, the most
likely to be selected is also the most likely to undergo a
spin-flip, namely the one which minimizes $\Delta
H(\mathbf{\sigma},j,\mathbf{J})$.

Let us now formalize how the dynamics  works. Our MC simulations
are made up of successive steps \cite{ABC09}:

- Being $i$ the newest updated spin/agent (at the very first step
$i$ is extracted randomly from the whole set of agents), we
consider the corresponding set of nearest-neighbors defined as
$\mathcal{N}_i=\{ i_1, i_2, ..., i_{\alpha_i}\}$; we possibly
consider also the subset $\tilde{\mathcal{N}}_i \subseteq
\mathcal{N}_i$ whose elements are nearest- neighbors of $i$ not
sharing the same orientation/opinion: $j \in \tilde{\mathcal{N}}_i
\Leftrightarrow j \in \mathcal{N}_i \wedge \sigma_i \sigma_j = -1$. Now, for
any $j \in \mathcal{N}_i$ we compute the cost function variation
$\Delta H (\mathbf{\sigma},j,\mathbf{J})$, which would result if the
flip $\sigma_j \rightarrow - \sigma_j$ occurred; notice that $\Delta H
(\mathbf{\sigma},j,\mathbf{J})$ involves not only the nearest-neighbors
of $i$.

- We calculate the probability of opinion-flip for all the nodes
in $\mathcal{N}_i$, hence obtaining
$p(\mathbf{\sigma},i_1,\mathbf{J}),p(\mathbf{\sigma},i_2,\mathbf{J}),...,p(\mathbf{\sigma},i_{\alpha_i},\mathbf{J})$,
where $p(\mathbf{\sigma},\sigma_j',\mathbf{J})$ (see eq.~\ref{eq:Glauber}),
is the probability that the current configuration $\mathbf{\sigma}$
changes due to a flip on the $j$-th site.

- We calculate the probability ${\cal P}^{\mathcal{S}}
(\mathbf{\sigma};i,j;\mathbf{J})$ that among all possible $\alpha_i$
opinion-flips considered just the $j$-th one is realized; this is
obtained by properly normalizing the $p(\mathbf{\sigma},j,\mathbf{J})$:
\begin{equation} \label{eq:probability}
{\cal P}^{\mathcal{S}} (\mathbf{\sigma};i,j;\mathbf{J}) =
\frac{p(\mathbf{\sigma},j,\mathbf{J})}{\displaystyle \sum_{k \in
{\mathcal{N}_i}} p(\mathbf{\sigma},k,\mathbf{J})}.
\end{equation}
We can possibly restrict the choice just to the set
$\tilde{\mathcal{N}}_i$, hence defining $\tilde{{\cal
P}}^{\mathcal{S}} (\mathbf{\sigma};i,j;\mathbf{J}) =
p(\mathbf{\sigma},j,\mathbf{J})/ \sum_{k \in {\tilde{\mathcal{N}}_i}}
p(\mathbf{\sigma},k,\mathbf{J})$.

- According to the normalized probability ${\cal P}^{\mathcal{S}}$
(see eq.~\ref{eq:probability}), we extract randomly the node $j
\in \mathcal{N}_i$ and realize the opinion flip $\sigma_{j}
\rightarrow - \sigma_{j}$.

- We set $j \equiv i$ and we iterate the procedure.

Finally, it should be underlined that in this dynamics detailed
balance is explicitly violated \cite{BBCV02,ABCV05}; indeed, its
purpose is not to recover a canonical Boltzmann equilibrium but
rather to model possible mechanism making the system evolve, and
ultimately, to describe, at an effective level, the statics
reached by a ``socially plausible'' dynamics for opinion spreading
\cite{ABC09}.

\subsection{Equilibrium behavior}\label{ssec:DDE}

The diffusive dynamics was shown to be able to lead the system
toward a well defined steady state and to recover the expected
phase transition, although the critical temperature revealed was
larger than the expected one \cite{BBCV02}. Such results were also
shown to be robust with respect to the the spin magnitude
\cite{ABCV05} and the underlying topology \cite{ABC09}. More
precisely, after a  suitable time lapse $t_0$  and for
sufficiently large systems, measurements of a (specific) physical
observable $X(\mathbf{\sigma},\beta)$ fluctuate around an average
value only depending on the external parameter $\beta$ and on the
geometry of the underlying structures (in particular, for diluted
systems, on $\alpha$). It was also verified that, for a system of
a given finite size $N$, the extent of such fluctuations scales as
$N^{-\frac{1}{2}}$ (see also \cite{BBCV02,ABCV05}), as
 indicated by standard statistical mechanics for a system in
 equilibrium. The estimate of the a given observable $\langle X(\beta)
 \rangle$ can therefore be obtained as an average over a suitable
 number of (uncorrelated) measurements performed when the system is
 reasonably close to the equilibrium regime.

Moreover, the final state obtained with the diffusive dynamics is
stable, well defined and, in particular, it does not depend on the
initial conditions, thus it displays all the properties of a
stationary state. Similar to what happens with the usual dynamics,
the relaxation time needed to drive the system sufficiently close
to the equilibrium situation is found to depend on the
temperature. More precisely, we experience the so called
`critical slowing down'': the closer $T$ to its critical value,
the longer the relaxation time.

\begin{figure}[tb]\bc
\includegraphics[height=60mm]{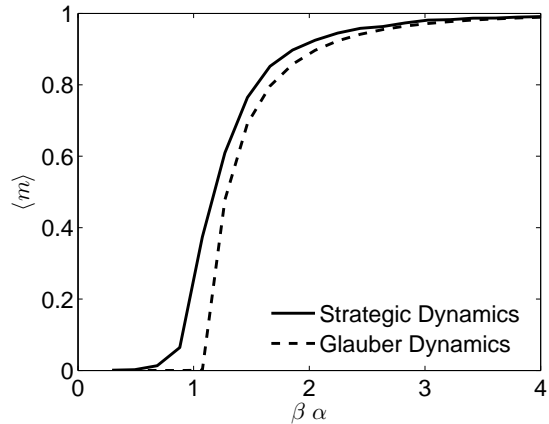}
\caption{\label{fig:p3} Critical behavior of the magnetization for
the two dynamics (diffusive and standard Glauber) as a function of
$\beta$ and fixed $\alpha=10$. The former dynamics gives rise to a critical point
higher with respect to the latter.} \ec
\end{figure}

In particular, it was evidenced that there exists a critical value
of the parameter $\beta_c^{\mathcal{S}}$ below which the system is
spontaneously ordered. However, $\beta_c^{\mathcal{S}}$ was found
to be appreciably smaller than the critical value
$\beta_c(\alpha)$ expected for the canonical Ising model on a
Erd\"{o}s-Renyi random graph. For example, we found
$\beta_c^{\mathcal{S}} \approx 0.07$ and $\beta_c^{\mathcal{S}}
\approx 0.016$ for $\bar{\alpha}=10$ and $\bar{\alpha}=45$,
respectively, versus $\beta_c(\bar{\alpha}) =
\tanh^{-1}(1/\bar{\alpha})$, yielding  $\beta_c(10) \approx 0.10$
and $\beta_c(45) =0.022$, (see Fig.$4$). Interestingly, it is not
possible describe the system subjected to the diffusive dynamics
by introducing an effective Hamiltonian obtained  from eq.
(\ref{ham}) by a trivial rescaling. In fact, calling $E$ the
numerical energies (to separate them from the analytical $e$), we
consider the dependence on the magnetization displayed by the
energies $E^{\mathcal{S}}(m)$ and $E(m)$, measured for system
evolving according to the diffusive dynamics and to a traditional
dynamics, respectively. As for the latter, from eq. (\ref{ham}) it
is easy to see that $E \propto m^2$. As for $E^{\mathcal{S}}(m)$,
we found that $E^{\mathcal{S}}<E$ for $0<m<1$, while
$E^{\mathcal{S}}=E$ for $m=0$ and $m=1$. This is compatible with a
power law behaviour $E^{\mathcal{S}} \sim m^ {2+\epsilon}$. In
order to obtain an estimate for $\epsilon$ we measured the ratio
$E^{\mathcal{S}}/E$ as a function of $m$; data are shown in the
log-log scale plot of Fig. $(5)$ and fitting procedures suggest
that $\epsilon \approx 0.15$. In the next section we will show
that this result can be interpreted as a consequence of an
effective $p$-agent interaction, being $p>2$.

\begin{figure}[tb]\bc
\includegraphics[width=70mm,height=50mm]{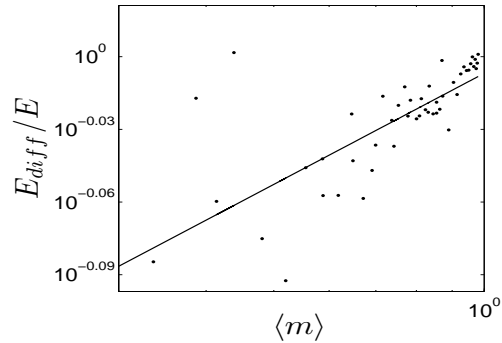}
\caption{\label{fig:p3} We extrapolate from $ m(\beta)$ and $E
(\beta)$ the plot $E(m)$ for both the dynamics. It is worth noting
that the diffusive dynamics deals to a curve living always
''below" the one obtained by Glauber dynamics results. This
strongly suggest the cooperation of $p>2$ spins each
interaction.}\ec
\end{figure}

\section{Statics of many body interactions}\label{tre}

So far, summing the discussion starting the sections $4$ and $5$,
we understood that interactions in social networks may involve
more than only couple exchanges. Consequently, corresponding to
this observation the need for a many-body imitative-behavior
Hamiltonian on a random graph appears. In this section, we
introduce such a $p$-spin model and study its thermodynamics.

\subsection{The diluted $p$-agent imitative behavior model}

We start exploiting the properties of a diluted even $p$-spin
imitative model: we restrict ourselves only to even values of $p$
for mathematical convenience (the investigation with the cavities
is much simpler), but, due to monotonicity of all the observables
in $p$, there is no need to see this as a real restriction (as
simulations on odd values of $p$ confirm and we will see later
on).
\newline
First of all, we define a suitable Hamiltonian acting on a
Erdos-Renyi random graph, with connectivity $\alpha$, made up by
$N$ agents $\sigma_{i}=\pm1, \ i \in [1,N]$.
\newline
Introducing $p$ families
$\{i_{\nu}^1\},\{i_{\nu}^2\},...,\{i_{\nu}^p\}$ of i.i.d. random
variables uniformly distributed on the previous interval, the
Hamiltonian is given by the following expression
\begin{equation}\label{hamp}
H_{N}(\sigma,\gamma(\alpha))=-\sum_{\nu=1}^{k_{\gamma (\alpha) N}}
\sigma_{i_{\nu}^1}\sigma_{i_{\nu}^2}...\sigma_{i_{\nu}^p}
\end{equation}
where, reflecting the underlying network, $k$ is a Poisson
distributed random variable with mean value $\gamma(\alpha) N$.
The relation among the coordination number $\alpha$ and $\gamma$
is $\gamma \propto \alpha^{p-1}$: this will be easily understood a
few line later by a normalization argument coupled with the high
connectivity limit of this mean field model.

The quenched expectation of the model is given by the composition
of the Poissonian average with the uniform one performed over the
families $\{i_{\nu}\}$
\begin{equation}
\textbf{E}[\cdot] = E_PE_i[\cdot] = \sum_{k=0}^{\infty}
\frac{e^{-\gamma(\alpha) N}(\gamma(\alpha)
N)^k}{k!N^p}\sum_{i_{\nu}^1....i_{\nu}^p}^{1,N}[\cdot].
\end{equation}

The Hamiltonian $(66)$, as written, has the advantage that it is
the sum of (a random number of) i.i.d.\ terms. To see the
connection to a more familiar Hamiltonian wrote in terms of
adjacency tensor $A_{i_1,...,i_p}$, we note that the
Poisson-distributed total number of bonds obeys $P_{\gamma
N}=\gamma N + O(\sqrt{N})$ for large $N$. As there are $N^p$
ordered spin p-plets $(i_1,...,i_p)$, each gets a bond with
probability $\sim \alpha/N$ for large $N$. The probabilities of
getting two, three (and so on) bonds scale as $1/N^2,1/N^3,\ldots$
so can be neglected. The probability of having a bond between any
unordered p-plet of spins is $p!$ as large, i.e.\ $2\alpha/N$ for
$p=2$.
\newline
It is possible to show that our version of the Hamiltonian in fact
is thermodynamically equivalent with the more familiar involving
the explicit adjacency tensor $A_{i_1,...,i_p}$, by recall at
first the latter model too:
\begin{equation}\label{parag}
-H_N(\sigma; k) \sim - \hat{H}_N(\sigma) = \sum_{i_1,...,i_p}^N
A_{i_1,...,i_p}\sigma_{i_1}...\sigma_{i_p},
\end{equation}
where $k$ is a Poisson variable with mean $\gamma N \sim
\alpha^{p-1} N$ and $A_{i_1,...,i_k}$ are all independent Poisson
variables of mean $\gamma/N^{p-1} \sim (\alpha / N)^{p-1}$.

Then, it is enough to consider the streaming of  the following
interpolating free energy (whose structure proves the statement a
priori by its thermodynamics meaning), depending on the real
parameter $t\in[0,1]$
$$
\phi(t) = \frac{\mathbb{E}}{N}\ln\sum_{\sigma}e^{\beta
\large(\sum_{\nu=1}^k \sigma_{i^1_{\nu}}...\sigma_{i^p_{\nu}} +
\sum_{i_1,...,i_p}^N A_{i_1,...,i_p}
\sigma_{i_1}...\sigma_{i_p}\large)},
$$
where $k$ is a Poisson random variable with mean $\gamma N t$ and
$A_{i_1,...,i_p}$ are random Poisson  variables of mean
$(1-t)\gamma/N^{p-1}$. Note that the two models alone are
recovered in the two extremals of the interpolation (for $t=0,1$).
By computing the $t$-derivative, we get
\begin{eqnarray}
\frac{1}{\gamma}\frac{d \phi(t)}{dt} &=& \mathbb{E}\ln(1+
\Omega(\sigma_{i_0^1}...\sigma_{i_0^p})\tanh(\beta))
\\ \nonumber &-& \frac{1}{N^{p}}\sum_{i_1,...,i_p}^N\ln(1+\Omega(\sigma_{i_1}...\sigma_{i_p})\tanh(\beta))=0,
\end{eqnarray}
where the label $0$ in $i_0^k$ stands for a new spin, born in the
derivative, accordingly to the Poisson property (\ref{Pp2}); as
the $i_0$'s are independent of the random site indices in the
$t$-dependent $\Omega$ measure, the equivalence is proved.

\bigskip

Following a statistical mechanics approach, we know that the
macroscopic behavior, versus the connectivity $\alpha$ and the
inverse temperature $\beta$, is described by the following free energy
density
\begin{eqnarray}
A(\alpha,\beta) &=& \lim_{N \to \infty} A_N(\alpha,\beta) \\
\nonumber &=& \lim_{N \to
\infty}\frac1N\textbf{E}\ln\sum_{\sigma}\exp(-\beta
H_{N}(\sigma,\gamma(\alpha))).\
\end{eqnarray}
The normalization constant can be checked by performing  the
expectation value of the cost function:
\begin{eqnarray}\nonumber
\textbf{E}[H] &=& -\gamma N m^p \\  \textbf{E}[H^2] -
\textbf{E}^2[H] &=& \gamma^2 N^2\Big[(q_{12}^p - m^p) +
O(\frac{1}{N})\Big],
\end{eqnarray}
by which it is easy to see that the model is well defined, in
particular it is linearly extensive in the volume. Then, in the
high connectivity limit each agent interacts with all the others
($\alpha \sim N$) and, in the thermodynamic limit, $\alpha \to
\infty$. Now, if $p=2$ the amount of couples in the summation
scales as $N(N-1)/2$ and, with $\gamma = 2\alpha$, a linear
divergence of $\alpha$ (desired to get a finite ratio $\alpha/N$
for each coupling) provides the right scaling; if $p=3$ the amount
of triples scales as $N(N-1)(N-2)/3!$ and, with $\gamma =
3!\alpha^2$, again we find the right connectivity behavior. The
generalization to every finite $p<N$ is straightforward.

\subsection{Properties of the random diluted  $p$-spin model}

Before starting our free energy analysis, we want to point out
also  the connection among this diluted version and the fully
connected counterpart (where each agents interact with the whole
community, not just a fraction as in the random network).
\newline
Let us remember that the Hamiltonian of the fully connected
$p$-spin model (FC) can be written as \cite{Bar08b} \be
H^{FC}_{N}(\sigma) = \frac{p!}{2 N^{p-1}}\sum_{1 \leq i_1 < ... <
i_p \leq N} \sigma_{i_1}\sigma_{i_2}...\sigma_{i_p}, \ee and let
us consider the  trial function $\hat{A}(t)$ defined as follows
\be \hat{A}(t) = \frac{1}{N}\mathbb{E}\ln \sum_{\sigma} \exp\Big[
\beta \sum_{\nu}^{P_{\gamma N
t}}\sigma_{i_{\nu}^1}\sigma_{i_{\nu}^2}...\sigma_{i_{\nu}^p} +
(1-t)\frac{\beta' N}{2}m^p \Big], \ee which interpolates between
the fully connected  $p$-spin model and the diluted one, such that
for $t=0$ only the fully connected survives, while the opposite
happens for $t=1$. Let us work out the derivative with respect to
$t$ to obtain \begin{eqnarray} \partial_t \hat{A}(t) &=&
(p-1)\alpha^{p-1}\ln\cosh(\beta) \\ \nonumber &-& (p-1)
\alpha^{p-1}\sum_{n}\frac{-1^n}{n}\theta^n \langle q_n^p\rangle -
\frac{\beta'}{2}\langle m^p \rangle, \end{eqnarray} by which we
see that the correct scaling, in order to recover the proper
infinite connectivity model, is obtained when $\alpha \to \infty$,
$\beta \to 0$ and $\beta' = 2(p-1) \alpha^{p-1}\tanh(\beta)$ is
held constant.

\begin{remark}
It is worth noting that in social modeling, usually, the role of
the temperature is left, or at least coupled together, to the
interaction strength $J$. As a consequence, in order to keep
$\beta'$ fixed, on different network dilution, the strength must
be rescaled accordingly to
$$
J = \tanh^{-1}\Big( \frac{\beta'}{2(p-1)\alpha^{p-1}} \Big),
$$
while, if present, an external field remains unchanged as it is a
one-body term, like $h\sum_i^N  \sigma_i$, unaffected by dilution.
\end{remark}
\begin{remark}
The dilute  $p$-spin model reduces to the fully connected one, in the
infinite connectivity limit, uniformly in the size of the system.
\end{remark}

\subsection{Properties of the free energy via the smooth cavity approach}

In this section we want to show some features of the free energy
corresponding to this model, which is investigated by extending
the previous method (the smooth cavity approach) to the many body
Hamiltonian.
\newline
As the generalization is simple and immediate to be achieved by
the reader, we skip the proofs in this section.
\newline
The starting point is always the representation theorem
\begin{theorem}\label{primolevi}
\textit{In the thermodynamic limit, the quenched pressure of the
even  $p$-spin diluted ferromagnetic model is given by the
following expression} \be A(\alpha,\beta) = \ln2
-\frac{\alpha}{p-1}\frac{d}{d\alpha}A(\alpha,\beta) +
\Psi(\alpha,\beta,t=1), \ee
\end{theorem}
where the cavity function $\Psi(t,\alpha,\beta)$ is introduced as
\begin{eqnarray}\label{cavity1}
&& \textbf{E}\Big[\ln\frac{\sum_{\{\sigma\}}
e^{\beta\sum_{\nu=1}^{k_{\tilde{\gamma}N}}
\sigma_{i_{\nu}^1}\sigma_{i_{\nu}^2}...\sigma_{i_{\nu}^p}}\;
e^{\beta\sum_{\nu=1}^{k_{2\tilde{\gamma}t}}
\sigma_{i_{\nu}^1}\sigma_{i_{\nu}^2}...\sigma_{i_{\nu}^{p-1}}}}
{\sum_{\{\sigma\}} e^{\beta\sum_{\nu=1}^{k_{\tilde{\gamma}N}}
\sigma_{i_{\nu}^1}\sigma_{i_{\nu}^2}...\sigma_{i_{\nu}^p}}}\Big]= \nonumber \\
&&  \textbf{E}\Big[\ln \frac{Z_{N,t}(\tilde{\gamma},\beta)}
{Z_{N}(\tilde{\gamma},\beta)}\Big] =
\Psi_N(\tilde{\gamma},\beta,t),
\end{eqnarray}
with $\lim_{N \to \infty}\tilde{\gamma}=\gamma$ as for
$\tilde{\alpha}$ and $\alpha$ in the two body model (see eq. $26$)
\be\label{cavity2} \Psi(\gamma,\beta,t) =
\lim_{N\rightarrow\infty}\Psi_N(\tilde{\gamma},\beta,t). \ee
\medskip
\newline
Thanks to the previous theorem, it is possible to figure out an
expression for the pressure by studying the properties of the
cavity function $\Psi(\alpha,\beta)$ and the connectivity shift
$\partial_{\alpha}A(\alpha,\beta)$.
\newline
Let us notice that
\begin{eqnarray}\label{Psit}
\frac{d}{d\alpha}A(\alpha,\beta) &=&
(p-1)\alpha^{p-2}\ln\cosh\beta - \\ \nonumber
 &-& (p-1)\alpha^{p-2}\sum_{n=1}^{\infty}\frac{(-1)^n}{n}\theta^n
\langle q_{1,...,n}^p \rangle,\\
\frac{d}{dt}\Psi(\tilde{\alpha},\beta,t) &=&
2\tilde{\alpha}^{p-1}\ln\cosh\beta - \\ \nonumber &-&
2\tilde{\alpha}^{p-1}\sum_{n=1}^{\infty}\frac{(-1)^n}{n}\theta^n
\langle q_{1,...,n}^{p-1} \rangle_{\tilde{\alpha},t}.
\end{eqnarray}

So we can understand the properties of the free energy by analyzing
the properties of the order parameters: magnetization and
overlaps, weighted in their extended Boltzmann state
$\tilde{\omega}_t$.
\newline
Further, as we expect the order parameters being able to describe
thermodynamics even in the true Boltzmann states $\omega, \Omega$,
accordingly to the earlier Definition $(2)$, we are going to
recall that {\em filled} order parameters (the ones involving even
numbers of replicas) are stochastically stable, or in other words,
are independent by the $t$-perturbation in the thermodynamic
limit, while the others, not filled, become filled, again in this
limit (such that for them $\omega_t\to \omega$ in the high $N$
limit and thermodynamics is recovered).

\begin{theorem}\label{ciccia}
In the thermodynamic limit and setting $t=1$ we have \be
\tilde{\omega}_{N,t}(\sigma_{i_1}\sigma_{i_2}...\sigma_{i_n}) =
\tilde{\omega}_{N+1}(\sigma_{i_1}\sigma_{i_2}...\sigma_{i_n}\sigma_{N+1}^n).
\ee
\end{theorem}
\begin{theorem}\label{saturabili}
Let $Q_{ab}$ be a not-filled monomial of the overlaps (this means
that $q_{ab}Q_{ab}$ is filled). We have \be
\lim_{N\rightarrow\infty}\lim_{t\rightarrow1} \langle Q_{ab}
\rangle_t = \langle q_{ab}Q_{ab} \rangle, \ee (examples:
\newline for $N \rightarrow \infty$ we get $\langle m_1 \rangle_t
\rightarrow \langle m_1^2 \rangle,\quad \langle q_{12} \rangle_t
\rightarrow \langle q_{12}^2 \rangle$).
\end{theorem}
\begin{theorem}\label{saturi}
In the $N\rightarrow\infty$ limit, the averages
$\langle\cdot\rangle$ of the filled polynomials are t-independent
in $\beta$ average.
\end{theorem}
\medskip

With the following definition \begin{eqnarray} \tilde{\beta} &=&
2(p-1)\tilde{\alpha}^{p-1}\theta \\ \nonumber &=&
2(p-1)\alpha^{p-1}\frac{N}{N+1}\theta \quad
\stackrel{N\rightarrow\infty}{\longrightarrow}
2(p-1)\alpha^{p-1}\theta = \beta', \end{eqnarray} we show the
streaming of replica functions, by which not filled multi-overlaps
can be expressed via filled ones.
\begin{proposition}\label{stream}
Let $F_s$ be a function of s replicas. Then the following
streaming equation holds
\begin{eqnarray}
\frac{\partial\langle F_s \rangle_{t,\tilde{\alpha}}}{\partial t}
&=& \tilde{\beta} \Big[\sum_{a=1}^s\langle F_s
m_a^{p-1}\rangle_{t,\tilde{\alpha}} - s \langle F_s
m_{s+1}^{p-1}\rangle_{t,\tilde{\alpha}}\Big] \quad
\\ \nonumber
&+& \tilde{\beta}\theta \Big[ \sum_{a<b}^{1,s}\langle F_s
q_{a,b}^{p-1} \rangle_{t,\tilde{\alpha}} - s\sum_{a=1}^s\langle
F_s q_{a,s+1}^{p-1}\rangle_{t,\tilde{\alpha}} \\ \nonumber &+&
\frac{s(s+1)}{2!}\langle F_s
q_{s+1,s+2}^{p-1}\rangle_{t,\tilde{\alpha}}\Big] + O(\theta^3).
\end{eqnarray}
\end{proposition}
\begin{remark}
We stress that, at the first two level of approximation presented
here, the streaming has has the structure of a $\theta$-weighted
linear sum of the Curie-Weiss streaming ($\theta^0$ term)
\cite{Bar08a} and the Sherrington-Kirkpatrick streaming
($\theta^1$ term) \cite{Bar06}, conferring a certain degree of
independence by the kind of quenched noise (frustration or
dilution) to mathematical structures of disordered systems.
\end{remark}

Overall the result we were looking for, a polynomial form of the
free energy, reads off as
\begin{eqnarray}\label{maino}
A(\alpha,\beta) &=& \ln2 \:+\: \alpha^{p-1}\ln\cosh\beta +\\
\nonumber &+& \frac{\beta'}{2}\Big(\beta'\langle m^{2(p-1)}\rangle
- \langle m^{p}\rangle\Big) + \\ \nonumber &+&
\frac{\beta'\theta}{4}\Big(\beta'\theta\langle
q_{12}^{2(p-1)}\rangle - \langle q_{12}^{p}\rangle\Big) +
O(\theta^5). \nonumber
\end{eqnarray}

Now, several conclusions can be addressed from the expression
(\ref{maino}):
\newline
\begin{remark}
At first let us note that, by constraining the interaction to be
pairwise, critical behavior should arise \cite{LL80}. Coherently,
we see that for $p=2$ we can write the free energy expansion as
$$A(\alpha,\beta)_{p=2} = \ln 2 + \alpha
\ln\cosh(\beta) - \frac{\beta'}{2}(1-\beta')\langle m^2 \rangle -
\frac{\beta' \theta}{4} \langle q_2^2 \rangle,$$ which coincides
with the one of the diluted two-body model (eq. $52$) and displays
criticality at $2\alpha\theta=1$, where the coefficient of the
second order term vanishes, in agreement with previous results
(sec. $4.3$).
\end{remark}
\begin{remark}
The free energy density of the fully connected $p$-spin model is
\cite{Bar08b} $A(\beta')= \ln 2 + \ln\cosh(\beta m^{p-1}) -
(\beta/2)m^p$, which coincides with the expansion (\ref{maino}) in
the limit of $\alpha \to \infty$ and $\beta \to 0$ with $\beta' =
2(p-1) \alpha^{p-1}\theta$ held constant.
\end{remark}
\begin{remark}
It is worth noting that the connectivity no longer plays a linear
role in contributing to the free energy density, as it does happen
for the diluted two body models \cite{ABC08,GT04}, but, in
complete generality as $p-1$. This is interesting in social
networks, where, for high values of coordination number it may be
interesting developing strategies with more than one exchange
\cite{NB01}.
\end{remark}

\subsection{Numerics}\label{sec:numerics}

We now analyze the system lastly described, from the numerical
point of view by performing extensive Monte Carlo simulations.
Within this approach it is more convenient to use the second
Hamiltonian introduced (see eq.(\ref{parag})):
\begin{equation}\label{my_hamiltonian}
\hat{H}_N(\sigma)=-\sum_{i_i}^{N}
\sigma_{i_1}\sum_{i_2<i_3<...<i_p=1}^{N} A_{i_1,...,i_p}
\sigma_{i_{2}}\sigma_{i_{3}}...\sigma_{i_{p}}.
\end{equation}
The product between the elements of the adjacency tensor ensures
that the $p-1$ spins considered in the second sum are joined by a
link with $i_1$.
\newline
The evolution of the magnetic system is realized by means of a
single spin flip dynamics based on the Metropolis algorithm
\cite{NB01}. At each time step a spin is randomly extracted and
updated whenever its coordination number is larger than $p-1$. For
$\alpha$ large enough (at least above the percolation threshold,
as  obviously it is the case for the results found previously) and
$p=3,4$ this condition is generally verified. The updating
procedure for a spin $\sigma_i$ works as follows: Firstly we
calculate the energy variation $\Delta e_i$ due to a possible spin
flip, which for $p=3$ and $p=4$ reads respectively
\begin{eqnarray}
\Delta e_i &=& 2 \sigma_i \sum_{j<k=1}^{N} A_{i,j}A_{i,k}
\sigma_{j}\sigma_{k},
\\
\Delta e_i &=& 2 \sigma_i \sum_{j<k<w=1}^{N} A_{i,j}A_{i,k}A_{i,w}
\sigma_{j}\sigma_{k}\sigma_{w}.
\end{eqnarray}
Now, if $\Delta e_i <0$, the spin-flip $\sigma_i \rightarrow -
\sigma_i$ is realized with probability $1$, otherwise it is
realized with probability $e^{-\beta \Delta e}$.

The case $p=3$ has been studied in details and some insight is
provided also for the case $p=4$, while  for $p=2$ we refer to
Sec. $(4.4)$. Our investigations concern two main issues:\\
- the existence of a phase transition and its nature\\
- the existence of a proper scaling for the temperature as the
parameter $\alpha$ is tuned.

As for the first point, we measured the so-called Binder cumulants
defined as follows:
\begin{equation}
G_N(T(\alpha)) \equiv 1 - \frac{\langle m^4 \rangle_N}{3\langle
m^2 \rangle_N^2},
\end{equation}
where $\langle \cdot \rangle_N$ indicates the average obtained for
a system of size $N$ \cite{Bin97}. The study of Binder cumulants
is particularly useful to locate and  catalogue the phase
transition. In fact, at any given connectivity (above the
percolation threshold), in the case of continuous phase
transitions, $G_N(T)$ takes a universal positive value at the
critical point $T_c$, namely all the curves obtained for different
system sizes $N$ cross each other. On the other hand, for a
first-order transition $G_N(T)$ exhibits a minimum at $T_{min}$,
whose magnitude diverges as $N$. Moreover, a crossing point at
$T_{cross}$ can be as well detected when curves pertaining to
different sizes $N$ are considered. Now, $T_{min}$ and $T_{cross}$
scale as $T_{min}-T_c \propto N^{-1}$ and $T_{cross}-T_c \propto
N^{-2}$, respectively.

\begin{figure}[tb]
\bc
\includegraphics[height=60mm]{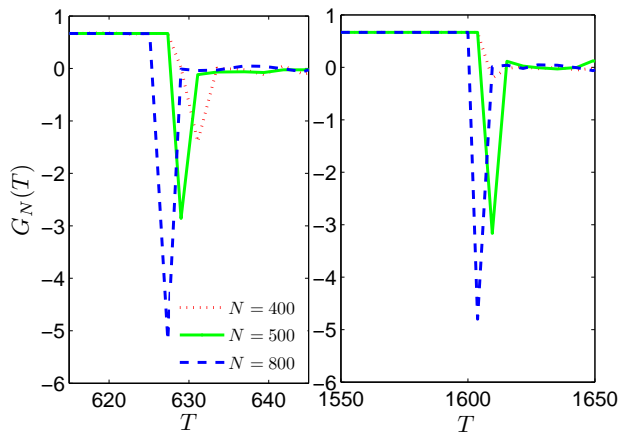}
\caption{\label{fig:Binder} Binder cumulants $G_L(T)$ for systems
of different size $N$, as shown in the legend, and connectivity
$\bar\alpha=50$ (left panel) and $\bar\alpha=80$ (right panel).}
\ec
\end{figure}

In Fig.~\ref{fig:Binder} we show data for $G_N(T)$ obtained for
systems of different sizes ($N=400$, $N=500$, and $N=800$) but
equal connectivity ($\alpha=50$ and $\alpha=80$, respectively) as
a function of the temperature $T$. The existence of a minimum is
clear and it occurs for $T \approx 625$ and $T \approx 1600$.
Similar results are found also for $p=4$ and they all highlight
the existence of a first-order phase transition at a temperature
which depends on the connectivity $\alpha$.

In order to deepen the role of connectivity in the evolution of
the system we measure the macroscopic observable $\langle m
\rangle$ and its (normalized) fluctuations $\langle m^2 \rangle -
\langle m \rangle^2$, studying their dependence on the temperature
$\beta$ and on the dilution $\alpha$. Data for different choices
of size and dilution are shown in Figure \ref{fig:p3}.

The profile of the magnetization, with an abrupt jump, and the
correspondent peak found for its fluctuations confirm the
existence of a first order phase transition at a well defined
temperature $T_c$ whose value depends on the dilution $\alpha$.
More precisely, by properly normalizing the temperature in
agreement with analytical results, namely $\tilde\beta \equiv
\beta \; \bar\alpha^{p-1}$ we found a very good collapse of all
the curves considered. Hence, we can confirm that the temperature
scales like $\alpha^{p-1}$.

\begin{figure}[tb]\bc
\includegraphics[height=60mm]{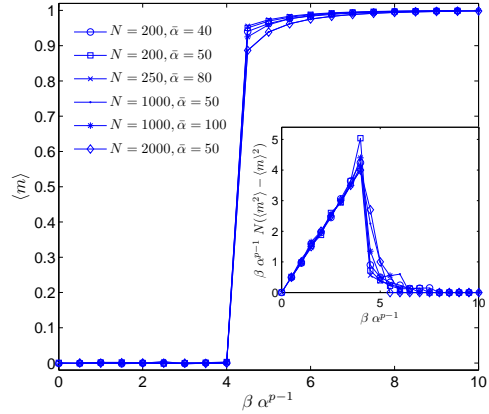}
\caption{\label{fig:p3}Magnetization (main figure) and its
normalized fluctuations (inset) for systems of different sizes and
different dilution as a function of $\beta \; \alpha^{p-1}$. The
collapse of all the curves provides a strong evidence for the
scaling of the temperature.} \ec
\end{figure}

\subsection{Diffusive dynamics revisited}
In the previous section we showed that the diffusive dynamics introduced give rise to a non-trivial thermodynamic which can not be explained by, say, a temperature rescaling. The crucial point is that such a dynamics intrinsically yields effective interactions which are more-than-two bodies.
This idea is supported by two important evidences (see \ref{ssec:DDE}):  a larger ``critical'' temperature $\beta_c^{\mathcal{S}}$ and the anomalous power-law behavior $E^{\mathcal{S}} \sim m^ {2.15}$,
which suggests the effective $p$
to be $2.15$. Now, while $p$ spans from two to infinity the critical temperature raises accordingly, hence we want to check that there
exist a suitable real value of $p$ that matches the critical
temperature found numerically and that is compatible with the
plots of $E^{\mathcal{S}}(m)$ and $E(m)$. When $p=2$ and close to criticality, the
temperature for the phase transition is given by $\beta_c =
\tanh^{-1}(1/2\alpha^{p-1})= \tanh^{-1}(1/2\alpha)$. For $p=2.15$ this expression becomes $\beta_c \sim
\tanh^{-1}(1/2\alpha^{p-1})= \tanh^{-1}(1/2\alpha^{1.15})$: The
ratio among the two expressions, when evaluated for $\alpha=10$ gets approximately $1.4$, in agreement with data depicted in Fig.~$4$.

Before concluding we notice that  for $p>2,
p\in\mathbb{N}$, ferromagnetic transitions are no longer critical
phenomena. At the critical line the magnetization is discontinuous and
a latent heat does exist. However, if $p$ is thought of as real, for
$p$ slightly bigger than two, as suggested by our data, the
``jump'' in the magnetization is expected to be small and to approach zero whenever $p\to2$.

\section{A simple application to trading in markets}

An appealing application of the whole theory developed concerns
trading among agents: Suppose we represent a market society only
with couple exchanges $(p=2)$, then there are just sellers and
buyers and they interact only pairwise. In this case if the buyer
$i$ has money ($\sigma_i=+1$) and the seller $j$ has the product
($\sigma_j=+1$), or if the buyer has no money and the seller has
no products ($\sigma_i=\sigma_j=-1$), the two merge their will and
the imitative cost function $(19)$ reaches the minimum. Otherwise,
if the seller has the product but the buyer has no money (or
viceversa), their two states are different $(\pm)$ and the cost
function is not minimized. In this scenario, the random graph
connects on average each agent to
 $\alpha$ acquaintances and this simply increases linearly the possibility
that each agent is satisfied. In fact, the higher the number of
``neighbors", the larger the possibility of trading.

When switching to the case $p=3$, other strategies are available:
for example the buyer may not have the money, but he may have a
valuable good which can be offered to a third agent, who takes it
and, in change, gives to the seller the money, so that the buyer
can obtain his target by using a barter-like approach. In this
case the two frustrated configurations from the previous sketch
can be avoided by multiplying by a factor $\sigma_k=-1$ given by
the third contributor $k$, or everything can remain the same of
course if the latter does not agree $(\sigma_k=+1)$.
Interestingly, we find that in this case $(p=3)$, the amount of
acquaintances one is in touch with (strictly speaking, the degree
of connectivity $\alpha$) does not contribute linearly as for
$p=2$, but quadratically: this seems to suggest that if a society
deals primarily with direct exchanges, no particular effort should
be done to connect people, while, if barter-like approaches are
allowed, then the more connected the society is, the larger is the
satisfaction reached on average by each agent in his specific
goal. The above scenario, intuitively, seems to match the contrast
among the classical barter-like approach of villages, where,
thanks to the small amount of citizens, their degree of reciprocal
knowledge is quite high and the money-mediated one of citizens in
big metropolis, where a real reciprocal knowledge is missing.

\section{Conclusions and Outlooks}

The idea to apply statistical mechanics methods to social and
economical sciences has appeared several years ago in the history
of science. The main drawback of the approach is the lack of a
proper measure criteria for the utility function. Although the
comparison may appear somehow risky one can say that the current
approach is at the same stage of the pre-thermodynamic epoch when
it wasn't clear at all that heat was a form of energy and both the
first and second principle of thermodynamics were still to be
identified. This parallel was indeed pointed out already by
Poincar\'e in reply to the Walras  theory of Economics and
Mechanics \cite{Wal09}. Poincar\'e said moreover that the lack of
ability to measure the utility function is not a severe obstacle
at a preliminary stage. What instead has proved to be a serious
problem in the development of mathematical approaches to economics
and social sciences has been the choice of axioms before
phenomenology had been studied properly and quantitative data had
been extensively analyzed (see \cite{Bou08}). The attempt that are
made nowadays, and we proposed an instance of statistical
mechanics nature, are to use mathematical models to mimic
microscopic realistic dynamics and reproduce to some extent the
typical macroscopic observed behavior. Further refinements of the
models will be necessary once the collection of data will start to
provide a fit with the free parameters introduced and hopefully
new principles and axioms will emerge after that stage.

\addcontentsline{toc}{chapter}{Bibliografia}
%
%
%

%
%

\end{document}